\documentclass[11pt,tightenlines,aps,amsmath,amssymb,nofootinbib,prd,shownopacs]{revtex4-2}
\usepackage{graphicx}
\usepackage{epstopdf}
\usepackage{latexsym}
\usepackage{amssymb}
\usepackage{amsmath}
\usepackage{color}
\usepackage{mathrsfs}
\usepackage{xparse}
\usepackage{float}
\usepackage{mathtools}
\usepackage[center]{subfigure}

\begin{document}
 \newcommand{\bq}{\begin{equation}}
 \newcommand{\eq}{\end{equation}}
 \newcommand{\bqn}{\begin{eqnarray}}
 \newcommand{\eqn}{\end{eqnarray}}
 \newcommand{\nb}{\nonumber}
 \newcommand{\cb}{\color{blue}}
    \newcommand{\cc}{\color{cyan}}
     \newcommand{\lb}{\label}
        \newcommand{\cm}{\color{magenta}}
\newcommand{\rc}{\rho^{\scriptscriptstyle{\mathrm{I}}}_c}
\newcommand{\rd}{\rho^{\scriptscriptstyle{\mathrm{II}}}_c}
\NewDocumentCommand{\evalat}{sO{\big}mm}{%
  \IfBooleanTF{#1}
   {\mleft. #3 \mright|_{#4}}
   {#3#2|_{#4}}%
}
\newcommand{\PRL}{Phys. Rev. Lett.}
\newcommand{\PL}{Phys. Lett.}
\newcommand{\PR}{Phys. Rev.}
\newcommand{\CQG}{Class. Quantum Grav.}
\newcommand{\md}{\mathrm{d}}

\title{On a novel relationship between shear and energy density at the bounce in non-singular Bianchi-I spacetimes}

\author{A. Meenakshi McNamara$^{1}$}
\email{mcnama20@purdue.edu}
\author{Sahil Saini$^{2}$}
\email{sahilsaiini@gjust.org}
\author{Parampreet Singh$^{3}$}
\email{psingh@lsu.edu}
\affiliation{$^{1}$ Department of Physics, Purdue University, West Lafayette, IN, 47907, USA\\
$^{2}$ Department of Physics, Guru Jambheshwar University of Science $\&$ Technology, Hisar, Haryana 125001, India\\
$^{3}$ Department of Physics and Astronomy, $\&$ Center for Computation and Technology, Louisiana State University, Baton Rouge, LA 70803, USA}

\begin{abstract}
In classical Bianchi-I spacetimes, underlying conditions for what dictates the singularity structure - whether it is anisotropic shear or energy density, can be easily determined from the generalized Friedmann equation. However, in non-singular bouncing anisotropic models these insights are difficult to obtain in the quantum gravity regime where the singularity is resolved at a non-vanishing mean volume which can be large compared to the Planck volume, depending on the initial conditions. Such non-singular models may also lack a generalized Friedmann equation making the task even more difficult.
We address this problem in an effective spacetime description of loop quantum cosmology (LQC) where energy density and anisotropic shear are universally bounded due to quantum geometry effects, but a generalized Friedmann equation has been difficult to derive due to the underlying complexity. Performing extensive numerical   simulations of effective Hamiltonian dynamics, we bring to light a surprising, seemingly universal  relationship between   energy density and  anisotropic shear at the bounce in LQC. For a variety of initial conditions for a massless scalar field, an inflationary potential, and two types of ekpyrotic potentials we find that the values of energy density and the anisotropic shear at the quantum bounce follow a novel parabolic relationship which reveals some surprising results about the anisotropic nature of the bounce, such as that the  maximum value of the anisotropic shear at the bounce is reached when the energy density reaches approximately half of its maximum allowed value.   The relationship we find can prove very useful  for developing our understanding of the degree of anisotropy of the bounce, isotropization of the post-bounce universe, and discovering the modified generalized Friedmann equation in Bianchi-I models with quantum gravity corrections.

\end{abstract}

\maketitle

\section{Introduction}

The Bianchi-I spacetime serves as a simple yet phenomenologically rich framework to understand the importance of anisotropies in the very early universe. For perfect fluids and matter content such as massive scalar fields with energy density $\rho$ and pressure $P$  which the satisfy equation of state $w = P/\rho < 1$, anisotropies dictate the dynamics close to the classical singularity and its geometrical shape, which is generally a cigar-like singularity.  For stiff matter, or a massless scalar field, the approach to singularity with some fine tuning can be isotropic (or point-like) but a cigar-like singularity remains prevalant \cite{jacobs,ellis-mm}. It is only when one considers a matter content allowing $w > 1$, such as in ekpyrotic scenarios \cite{turok-steinhardt,turok-steinhardt1,Buchbinder:2007ad,CaiBranden2012}, that a more probable isotropization of the spacetime structure near the singularity can happen.  These insights are easily accessible via the availability of a generalized Friedmann equation in the classical Bianchi-I spacetime which relates the expansion scalar to anisotropic shear and energy density \cite{ellis}.  But the classical theory is incomplete due to the presence of singularities and it has been long expected that the latter would be resolved by quantum gravity. An important question is how does the new physics responsible for singularity resolution change the interplay between anisotropic shear and energy density? To resolve the singularity, such a modified theory of gravity must tame the divergences in anisotropic shear and energy density, and as a result the spacetime curvature. Given that a generalized Friedmann equation with quantum gravity corrections may no longer be available in such a setting and the singularity may be resolved via a bounce at a non-vanishing mean volume, insights from classical theory have a limited validity. Certainly one can not expect that the geometrical nature of the bounce would be the same as that of the to-be singularity in the classical theory for a given equation of state. Another way to state this issue is to ask that if in a  non-singular anisotropic model, a bounce occurs at a non-zero scale factor for some given matter content, how anisotropic or isotropic is the bounce?

The goal of this paper is to answer these questions in the setting of loop quantum cosmology (LQC) \cite{as-status} which is based on using  techniques of non-perturbative quantization of gravity as understood in loop quantum gravity (LQG).  The main result of LQC is the resolution of the big bang singularity which is replaced by a quantum bounce in the Planck regime \cite{aps1,aps2,aps3}, a prediction which has been tested for a wide variety of states using high performance computing \cite{numlsu-1,numlsu-2,numlsu-3, DienerJoeMegSingh2017,Singh:2018rwa}. For a spatially-flat isotropic model sourced with a massless scalar field, an exactly soluble model shows that expectation values of the energy density operator have a universal maximum value fixed by the area gap in quantum geometry \cite{slqc}. Further, the probability of bounce in consistent histories formalism turns out to be unity \cite{craig-singh}. A fair amount of work has been done to understand the Bianchi-I model in LQC in recent years which can be summarized as follows. Loop quantization of the Bianchi-I model results in a quantum difference equation \cite{AshtekerWE2009, Martin-Benito:2009xaf} as in isotropic models and it has been shown through extensive numerical simulations that the quantum dynamics of the Bianchi-I model can be extremely well captured by an effective continuum spacetime description  \cite{DienerJoeMegSingh2017}. Moreover, it has been shown that the energy density and the shear scalar are bounded from above in this effective spacetime \cite{Corichi:2009pp, GuptSingh2012}. An analysis of singularities shows that the effective Bianchi-I spacetime is generically devoid of any strong curvature spacetime singularities \cite{PS2012}, a result also valid for Bianchi-II \cite{ss1} and Bianchi-IX spacetimes \cite{ss2}. Numerical investigations show that the quantum bounce is accompanied by certain specific Kasner transitions in the effective Bianchi-I spacetime \cite{GuptPSKasner2012,flat-kasner,EwingKasner2018}. It has been further demonstrated that a viable non-singular inflationary model can be constructed starting from highly anisotropic initial conditions in Bianchi-I effective spacetime \cite{GuptInflationB-I, barrau2}.  With the assumption that anisotropies are small at the bounce, imprint on cosmological perturbations in the effective Bianchi-I spacetime for inflationary models has been studied \cite{AgulloCMBinB-I}. The Bianchi-I model in LQC has also been studied for ekpyrotic potentials and it has been shown that singularity can be resolved with bounded anisotropies \cite{CalitauPSKV2009}. Finally, ongoing work on Gowdy models aims to include Fock quantized inhomogenities in a loopy Bianchi-I setting (see \cite{gowdy} for a review).

Despite this progress for understanding quantum geometric effects in Bianchi-I model in LQC, some important questions have remained open. Let us note that in LQC it is the gravitational part of the Hamiltonian constraint which is modified, and in isotropic models of LQC, quantum gravity effects from geometrical part of the effective Hamiltonian can be translated to the modifications of energy density to result in a modified Friedmann equation  which captures quantum evolution and non-singular bounce quite accurately \cite{aps2,Singh:2006sg}. But, this exercise to transfer quantum geometric modifications in gravitational side to matter side quickly becomes complicated in more general models. As discussed in Sec. III, such an equation has been difficult to derive in anisotropic models from the effective Hamiltonian constraint due to  underlying complexity. As a result, while the availability of a generalized Friedmann equation in the classical regime aids in understanding of the interplay between the energy density and anisotropic shear, no such equation is yet known for the effective spacetimes in anisotropic models in LQC. As a result, there are few analytical insights on the way quantum geometry affects energy density and shear in the Planck regime and any relationship between them is cloaked. So far it is only known that the anisotropic shear scalar as well as the energy density of  the matter content are universally bounded in the Planck regime but in the anisotropic model the energy density and shear in general never reach these values because they both contribute to the spacetime curvature. %
The complicated form of the equations of motion in the effective spacetime do not lend themselves easily to a qualitative analysis. Taking cue from the fact that numerical simulations have in the recent past been key to uncovering the rich phenomenology of the effective Bianchi-I spacetime in the quantum regime, we employ extensive numerical simulations in this work and uncover a surprisingly simple relationship between the energy density and the shear scalar at the bounce.
%We hope that this adds an important piece to the understanding of the dynamics of the effective Bianchi-I model in the quantum regime.

%Since it is important to consider quantum geometry effects in the very early universe, we consider here the effective Bianchi-I spacetime in the framework of loop quantum cosmology (LQC).

We perform a large number of simulations with initial conditions chosen randomly from a given range, we note the value of the energy density and shear scalar at the bounce in each case. Such a computationally heavy approach is necessitated in this case as the overall trend cannot be discerned from the data of only a few simulations. A set of more than 150 such simulations is performed for each different scalar field potential we consider in this manuscript. The fields considered here are the massless scalar field, the massive scalar field with quadratic potential and two types of ekpyrotic fields. The quadratic potential is the simplest case of a matter field with a potential and is also extensively used in studies involving inflation. The ekpyrotic fields, though originally introduced in the context of string theory \cite{turok-steinhardt, turok-steinhardt1}, are often used in matter-ekpyrotic alternatives to inflation where a dust-dominated contracting regime produces a scale invariant spectrum while the ekpyrotic field becomes important near the bounce to guard against the anisotropic BKL instability \cite{turok-steinhardt1,CaiBranden2012,CalitauPSKV2009,CaiEwingEkp2014,HaroCai2015,LiSainiSingh2021}. We consider two different ekpyrotic potentials, the original form introduced in \cite{turok-steinhardt} and a closely related form often employed in matter-bounce scenarios \cite{CaiBranden2012}.

Examining the data for various distinct choices of the matter fields, we find that there exists a so far unknown parabolic relationship between the energy density and the shear scalar at the bounce for each type of matter field. Availability of such a relationship is a priori unobvious if one considers only a few simulations as is generally the case so far in LQC. And, this simple parabolic relationship at the bounce is in a stark contrast to the complicated expressions for the equations of motion and other physical quantities in the effective dynamics which would have been very difficult to guess from the dynamical equations of effective spacetime without performing simulations with extensive initial conditions.
Moreover, the fit parameters seem to depend weakly on the choice of the matter field, indicating that the parabolic relationship at the bounce may be a feature of the underlying effective dynamics of the Bianchi-I model in LQC. We discuss some implications of this parabolic relationship at the bounce on the generalized effective Friedmann equation. An unexpected finding is that the anisotropic shear at the bounce $\sigma^2_b$ takes its maximum allowed value in LQC approximately when energy density at the bounce $\rho_b$ reaches half of its maximum value. We also find that earlier approximations to the generalized effective Friedmann equation \cite{ChiouKV2007}, derived with the assumptions of low anisotropies, are disfavored by our numerical results. This is expected as we find from our numerical simulations that on average, the anisotropic shear dominates the isotropic energy density at the bounce. We provide a general form for the generalized effective Friedmann equation which is consistent with our results. We note that much further exploration is required to probe the nature of dynamics in the quantum regime, and hope that our result can provide a crucial stepping stone in that direction.
While our analysis is restricted to LQC, our methodology can be easily applied to other bouncing models and we hope that results presented here would yield insights on the degree of anisotropy of bounce in a more general setting.

This manuscript is organized as follows. Section II provides a brief summary of the classical dynamics of the Bianchi-I model using Ashtekar-Barbero variables. This is followed by Section III discussing the essential features of the effective dynamics of the Bianchi-I model in LQC, including the effective equations of motion we use for numerical simulations. The results of numerical simulations with a variety of initial conditions and different matter fields are described in Section IV. Note that in this section all values are in Planck units. In Sec. V we use results from numerical simulations to develop some insights on the form of the modified Friedmann equation in Bianchi-I model in LQC.  We summarize our conclusions in Section VI.

\section{Classical setting of the Bianchi-I spacetime}
% \renewcommand{\theequation}{2.\arabic{equation}}\setcounter{equation}{0}

% \subsection{The dynamical equations}
In this section we discuss some of the classical aspects of the Bianchi-I spacetime  in the canonical framework using Ashtekar-Barbero variables. We consider the diagonal Bianchi-I spacetime with foliation $\Sigma \times \mathbb{R}$ where $\Sigma$ is a spatially-flat hyperspace with $\mathbb{R}^3$ topology. Taking the lapse as $N=1$, the line element of the spacetime in comoving coordinates is given by
\begin{equation}
    \md s^2=- \md t^2+a_1^2 \md x^2 + a_2^2 \md y^2 + a_3^2 \md z^2,
\end{equation}
where $a_1(t)$, $a_2(t)$ and $a_3(t)$ are the directional scale factors. Due to the underlying symmetries of the spacetime, once we impose the Gauss and the spatial-diffeomorphism constraints, the Ashtekar-Barbero connection $A^i_a$ and the triads $E^a_i$ have only one independent component per spatial direction. The symmetry reduced connections $c^i$ and triads $p_j$ satisfy the following Poisson brackets
\begin{equation}\label{eq:pb}
    \lbrace c^i,p_j \rbrace = 8\pi G \gamma \delta^i_j,
\end{equation}
where $\gamma$ is the Barbero-Immirizi parameter. For numerical simulations we will fix its value as $\gamma=0.2375$ using black hole thermodynamics in LQG. In terms of the symmetry reduced Ashtekar-Barbero variables, the classical Hamiltonian constraint for Bianchi-I spacetime with a matter content minimally coupled to gravity is given by\footnote{The directional triad can have positive and negative orientation. We assume the orientation to be positive.}
\begin{equation}
    \mathcal{H_\mathrm{cl}}=-\frac{1}{8\pi G \gamma^2 v} \left(c_1 p_1 c_2 p_2 + \mathrm{cyclic~terms} \right) + \mathcal{H_\mathrm{m}} .
\end{equation}
Here $\mathcal{H_\mathrm{m}}$ denotes the matter Hamiltonian and $v=\sqrt{p_1 p_2 p_3}=a_1 a_2 a_3$ is the physical volume of a unit comoving cell.\footnote{In non-compact models in LQC, one introduces a fiducial cell to define symplectic structure whose coordinate lengths enter the relation between triads and scale factors. We assume the coordinate lengths of this fiducial cell to be unity.}  With all other constraints fixed, and the lapse equal to unity, the Hamiltonian is given by the above expression.

While the triads are kinematically related to the directional scale factors as
\begin{equation}
    p_1=a_2 a_3, \quad p_2=a_1 a_3, \quad p_3=a_1 a_2,
\end{equation}
the directional connection components are determined via Hamilton's equations as $c^i = \gamma \dot a_i$. The Hamilton's equations can be computed as
\begin{equation}
\dot p_i = \{p_i,  \mathcal{H_\mathrm{cl}}\} = - 8 \pi G \gamma \frac{\partial  \mathcal{H_\mathrm{cl}}}{\partial c_i}, ~~~   \dot c_i = \{c_i,  \mathcal{H_\mathrm{cl}}\} =  8 \pi G \gamma \frac{\partial  \mathcal{H_\mathrm{cl}}}{\partial p_i}   ~.
\end{equation}

In this work, we consider the case where matter is a perfect fluid with equation of state $w = P/\rho$ where $\rho = {\cal H}_{\mathrm{m}}/v$ and $P = - \partial {\cal H}_{\mathrm{m}}/\partial v$ denote the energy density and pressure of the perfect fluid respectively. The Hamilton's equations of motion yield the following evolution equations for the directional Hubble rates,
\begin{equation}\label{eq:H1H2}
    H_1 H_2+H_2 H_3 + H_3 H_1 = 8\pi G \rho,
\end{equation}
and similarly
\begin{equation}
    \dot H_2 + \dot H_3 + H_1^2 + H_2^2 + H_3^2 = - 8 \pi G P, \quad \text{and its cyclic permutations},
\end{equation}
 The directional Hubble rates $H_i$ are given by
\begin{equation}\label{eq:Hubble-triad}
H_1=\frac{\dot a_1}{a_1} = \frac{1}{2} \left(\frac{\dot p_2}{p_2} + \frac{\dot p_3}{p_3} - \frac{\dot p_1}{p_1} \right)
\end{equation}
and similarly for $H_2$ and $H_3$.

An important measure of the anisotropy of the spacetime is the shear scalar, which is defined as $\sigma^2=\sigma^{ij}\sigma_{ij}$ where $\sigma_{ij}$ is the anisotropic shear tensor. The anisotropic shear is the traceless part of the expansion tensor $\theta_{\alpha \beta}$ given by the covariant derivative of the unit fluid velocity, and takes a diagonal form $\sigma_{ij}=diag(\sigma_1,\sigma_2,\sigma_3)$ for the diagonal Bianchi-I spacetime. The diagonal components are given as $\sigma_i=H_i-\frac{1}{3}\theta$ where $\theta$ is the trace of the expansion tensor
\begin{equation}
\theta = H_1 + H_2 + H_3 = 3 H
\end{equation}
and $H$ is the mean Hubble rate. The shear scalar is given by
\begin{equation}\label{eq:shear}
    \sigma^2=\sum_{i} \sigma_i^2 = \frac{1}{3} \left((H_1-H_2)^2+(H_2-H_3)^2+(H_3-H_1)^2 \right).
\end{equation}
Note that above relations of expansion scalar and shear to Hubble rates are kinematical relations
which do not depend on whether the underlying theory is GR or LQC.

The dynamical equations for matter with vanishing anisotropic stress imply that $(H_i-H_j)= \alpha_{ij}/a^3$ where $a \equiv (a_1 a_2 a_3)^{1/3}$ is the mean scale factor and $\alpha_{ij}$ are constants. The shear scalar can thus be written as,
\begin{equation}
    \sigma^2= 6\, \frac{\Sigma^2}{a^6},
\end{equation}
where $\Sigma^2=(\alpha_{12}^2+\alpha_{23}^2+\alpha_{31}^2)/18$
is a constant of motion in the classical theory. Using the shear scalar and the mean Hubble rate $H$, evolution equations for directional Hubble rates can be recasted as the following generalized Friedmann and Raychaudhuri equations:
\begin{eqnarray}\label{Friedmann}
    H^2 &=& \frac{8\pi G \rho}{3}+\frac{\sigma^2}{6}, \label{Friedmann} \\
    \dot H &=& -\frac{1}{2}\left( 8\pi G (\rho+P)+ \frac{3\sigma^2}{2}\right).\label{Raychaudhuri} ~
\end{eqnarray}
The generalized Friedmann equation shows the way anisotropic shear can be understood at the same level of matter energy density to understand the expansion rate of the universe. Note that this involved expressing directional Hubble rates appearing on the
left hand side of \eqref{eq:H1H2} in terms of the anisotropic shear on the right hand side of \eqref{Friedmann}. While classically this is possible due to a simple form the Hamiltonian constraint or equivalentaly \eqref{eq:H1H2}, it is not guaranteed to occur when the Hamiltonian constraint gets on-trivial modifications as in LQC.

 The availability of the generalized Friedmann equation aids greatly in analyzing and understanding the dynamics of the universe. For example, let us note that the energy density and pressure satisfy the conservation law:
\begin{equation}
\dot \rho + 3 H (\rho + P) = 0
\end{equation}
which implies $\rho \propto a^{-3(1 + w)}$ for a constant equation of state. Since the shear scalar scales as $a^{-6}$, it grows faster than any perfect fluid with equation of state $w < 1$ on approach to singularity. If one considers a stiff matter which has equation of state $w = 1$, or equivalently a massless scalar, there is a competition between the anisotropic shear and the energy density in dictating the dynamics. Yet, the structure of the spacetime near the singularity is not isotropic but generally anisotropic in the sense that the approach to the singularity favors a cigar-like singularity \cite{jacobs}.  Thus,  the contribution of anisotropies in the generalized Friedmann equation is likely to dominate over the energy density in the regime when volume is small and equation of state is $w \leq 1$. For this reason, anisotropies are expected to play an important role in the very early universe. However, if one has matter content which satisfies $w > 1$, such as for a scalar field in a negative potential in ekpyrotic scenarios \cite{turok-steinhardt, turok-steinhardt1, CaiBranden2012}, then we can easily see from \eqref{Friedmann} that the energy density plays a dominant role near the singularities and effects of anisotropic shear can be dampened under right conditions.

\section{Effective dynamics of the Bianchi-I model in LQC}
LQC applies the concepts and methods from LQG to symmetry reduced cosmological models. The quantization in this approach is based on the holonomies of the connection and the fluxes of the triads as the basic variables. Often, a continuum effective description can be found in LQC, which has been shown to faithfully approximate the quantum dynamics of a wide variety of quantum states for various models such as the isotropic model \cite{numlsu-1,numlsu-2,numlsu-3}, and in particular for Bianchi-I model \cite{DienerJoeMegSingh2017, Singh:2018rwa}. In this work, we rely on this effective description of the loop quantized Bianchi-I model. The effective Hamiltonian for the Bianchi-I model for lapse $N=1$ is given by \cite{AshtekerWE2009,ChiouKV2007,CalitauPSKV2009}
\begin{eqnarray}
\label{hamiltonian}
    \mathcal{H}&=&-\frac{1}{8\pi G \gamma^2 v} \left( \frac{\sin(\bar\mu_1 c_1)}{\bar\mu_1} \frac{\sin(\bar\mu_1 c_2)}{\bar\mu_2} p_1 p_2 + \mathrm{cyclic ~terms} \right) + \mathcal{H_\mathrm{m}}\\ \nonumber
    &=&-\frac{v}{8\pi G \gamma^2 \lambda^2} \left(\sin(\bar\mu_1 c_1) \sin(\bar\mu_1 c_2) + \mathrm{cyclic ~terms} \right) + \mathcal{H_\mathrm{m}},
\end{eqnarray}
where $ \lambda = \sqrt{\Delta} $ with $ \Delta = 4 \sqrt{3} \pi \gamma \ell_{\mathrm{Pl}}^2 $. The holonomy edge lengths are given by
\begin{equation}\label{mu_term}
    \bar \mu_1=\lambda \sqrt{\frac{p_1}{p_2 p_3}}, ~\bar \mu_2=\lambda \sqrt{\frac{p_2}{p_3 p_1}}, ~\bar \mu_3=\lambda \sqrt{\frac{p_3}{p_1 p_2}}  .
\end{equation}
While modified generalized Friedmann and Raychaudhuri equations are not known for the Bianchi-I model in LQC, the effective Hamiltonian given in \eqref{hamiltonian} determines the dynamics of the spacetime through the triad and connection variables.

The equations of motion can be determined using the fundamental Poisson bracket \eqref{eq:pb} as follows:
\begin{align}\label{triad_flow}
    \dot{p_1} =&\frac{p_1}{\gamma\lambda}\left(\sin(\overline{\mu}_2c_2) + \sin(\overline{\mu}_3c_3)\right)\cos(\overline{\mu}_1c_1), \\
    \dot{c_1} =& \frac{V}{2\gamma \lambda^2 p_1}\Big[ c_2\overline{\mu}_2\cos(\overline{\mu}_2c_2)\left(\sin(\overline{\mu}_3c_3) + \sin(\overline{\mu}_1c_1)\right) +
    c_3\overline{\mu}_3\cos(\overline{\mu}_3c_3)\left(\sin(\overline{\mu}_1c_1) + \sin(\overline{\mu}_2c_2)\right) \nonumber \\
    & ~~~~~~~~~~ - c_1\overline{\mu}_1\cos(\overline{\mu}_1c_1)\left(\sin(\overline{\mu}_2c_2) + \sin(\overline{\mu}_3c_3)\right) \nonumber\\
    & ~~~~~~~~~~ - \left(\sin(\overline{\mu}_1c_1)\sin(\overline{\mu}_2c_2) + \sin(\overline{\mu}_2c_2)\sin(\overline{\mu}_3c_3) + \sin(\overline{\mu}_3c_3)\sin(\overline{\mu}_1c_1)\right) \Big] \nonumber\\
    & ~~~~~~~~~~+ 8\pi G\gamma \frac{\partial \mathcal{H}_\mathrm{m}}{\partial p_1},
\end{align}
with similar equations for the other triad and connection components. From the vanishing of the Hamiltonian constraint, we can also obtain the following expression for the energy density:
\begin{align}
    \rho &= \frac{1}{8\pi G\gamma^2\lambda^2}\left(\sin(\overline{\mu}_1c_1)\sin(\overline{\mu}_2c_2) + \sin(\overline{\mu}_2c_2)\sin(\overline{\mu}_3c_3) + \sin(\overline{\mu}_1c_1)\sin(\overline{\mu}_3c_3) \right),
\end{align}
which is bounded as opposed to the divergent behavior of $\rho$ in the classical case. The upper bound on energy density turns out to be
\begin{align}
    \rho \leq \rho_{\mathrm{max}} = \frac{3}{8\pi G\gamma^2\lambda^2} \approx 0.41\rho_{\mathrm{Pl}}
\end{align}
which is the same as the upper bound on energy density in LQC of the isotropic model \cite{aps3,Singh:2006sg}.

From the equations of motion for triads, and using \eqref{eq:Hubble-triad} we can obtain the directional Hubble rates in effective dynamics as:
\begin{equation}
    H_1 = \frac{\dot{a_1}}{a_1} = \frac{1}{2\gamma\lambda}\left(\sin(\overline{\mu}_1c_1 - \overline{\mu}_2c_2) + \sin(\overline{\mu}_1c_1-\overline{\mu}_3c_3) + \sin(\overline{\mu}_2c_2 + \overline{\mu}_3c_3) \right)
\end{equation}
with similar equations for $H_2$ and $H_3$. It is clear from this equation that unlike in the classical case, the Hubble rates are bounded in the effective LQC dynamics. We can use these to evaluate the shear scalar:
\begin{align}
    \sigma^2 =& \frac{1}{3}\left((H_1-H_2)^2 + (H_2-H_3)^2 + (H_3-H_1)^2\right) \\\nonumber
    =& \frac{1}{3\gamma^2\lambda^2}\Big[\left(\cos(\overline{\mu}_3c_3)(\sin(\overline{\mu}_1c_1) + \sin(\overline{\mu}_2c_2)) - \cos(\overline{\mu}_2c_2)(\sin(\overline{\mu}_1c_1) + \sin(\overline{\mu}_3c_3))\right)^2\\\nonumber
    & + \left(\cos(\overline{\mu}_3c_3)(\sin(\overline{\mu}_1c_1) + \sin(\overline{\mu}_2c_2)) - \cos(\overline{\mu}_1c_1)(\sin(\overline{\mu}_2c_2) + \sin(\overline{\mu}_3c_3))\right)^2\\\nonumber
    & + \left(\cos(\overline{\mu}_2c_2)(\sin(\overline{\mu}_3c_3) + \sin(\overline{\mu}_2c_2)) - \cos(\overline{\mu}_1c_1)(\sin(\overline{\mu}_2c_2) + \sin(\overline{\mu}_3c_3))\right)^2\Big].
\end{align}
It is important to note that unlike classical theory, the shear scalar is bounded by a universal value in LQC \cite{Corichi:2009pp}.
%We note that $\sigma^2$ and is now bounded.
The maximum anistropic shear possible in the LQC effective dynamics is
\begin{equation}
    \sigma^2 \leq \sigma_{\mathrm{max}}^2 = \frac{10.125}{3\gamma^2\lambda^2}\approx 11.57 l_{\mathrm{Pl}}^{-2}.
\end{equation}
The boundedness of the energy density, Hubble rates and shear scalar are strong indicators that classical singularities are absent from the effective spacetime. This turns out to be true and it can be shown that for arbitrary matter, strong curvature singularities are absent in the effective spacetime of Bianchi-I model in LQC \cite{PS2012}. Moreover, numerical simulations show that the singularities are replaced by a bounce which in the case of Bianchi-I spacetimes is accompanied by Kasner transitions in the geometry of the spacetime \cite{GuptPSKasner2012}.

While the existence of the bounce is a generic feature of Bianchi-I LQC as in the isotropic models, there are some important differences due to anisotropies. Unlike in the isotropic case, the interplay between the energy density and the anistropic shear can result in neither $\rho$ nor $\sigma^2$ obtaining their maximum values $\rho_{\mathrm{max}}$ and $\sigma_{\mathrm{max}}^2$ at the bounce. The exact values taken by these physical quantities varies from one one set of initial data to another with a priori no simple governing relation between them. As a result, the question of how anisotropic is the bounce under a generic set of initial conditions and the choice of matter has been a difficult question to answer. This can be answered by understanding the relationship of energy density and shear scalar such as the case in the classical model where a generalized Friedmann equation is available.
However, unlike the isotropic model in LQC, a modified generalized Friedmann equation has not been possible to obtain for the Bianchi-I model in LQC due to the very complicated expressions for the equations of motion as shown above and the form of the Hamiltonian constraint. The lack of a generalized Friedmann equation is a hindrance to understand details of the dynamics, in particular to gain insights on the way energy density and anisotropic shear play roles in singularity resolution. Apart from noting some important generalities, such as, the energy density, the Hubble rates and the shear scalar are generically bounded during evolution and the singularities are resolved, it is not possible to gain physical intuition analytically. Hence, the only possible resort at present is to perform numerical simulations using Hamilton's equations.

We note that any modified generalized Friedmann equation obtained for Bianchi-I model must reduce to equation \eqref{Friedmann} in the classical limit. Thus it must depend on both $\rho$ and $\sigma^2$. An attempt has been made earlier in \cite{ChiouKV2007} to obtain a modified generalized Friedmann equation for Bianchi-I model in LQC in the limit of small anisotropic shear, where an expression quadratic in both $\rho$ and $\sigma^2$ is obtained. However, as we have noted in the previous section, anisotropies are likely to play a key role near the classical singularities and are likely to be the dominant contributor to the dynamics near the quantum bounce where departures from classical behavior are expected to become significant. Thus, the approximation of weak anisotropic shear is expected to be invalid in this regime, which is the main regime of interest, for which the modified generalized Friedmann equation is lacking.\footnote{It is possible to recast the Hamiltonian constraint in the form of a ``modified Friedmann equation'' albeit by changing the kinematical definition of the shear scalar. For an attempt in this direction, see \cite{barrau}, where one defines a ``quantum shear''  which has no transparent relation to Hubble rates and also depends on matter Hamiltonian. As a result, such an equation can not capture the relation between energy density and anistropic shear as defined through differences in directional Hubble rates \eqref{eq:shear}.}
In absence of a modified Friedmann equation in Bianchi-I models in LQC, the only way to understand the relationship between energy density and shear at the bounce is via numerical simulations using Hamilton's equations. This is the goal of our numerical studies in this paper
where using extensive numerical simulations we uncover surprisingly simple and generic relationship between the energy density and the shear scalar at the bounce.

\section{Numerical Analysis of Bianchi-I spacetime in LQC}

In order to explore the relationship between $\rho$ and $\sigma^2$ at the bounce, denoted as $\rho_b$ and $\sigma^2_b$ respectively, we perform extensive  numerical simulations using the effective equations of motion obtained in the previous section.  We also consider a variety of different matter content to understand the genericity of relationship between energy density and shear at the bounce. Since the anisotropic shear always dominates evolution unless the equation of state of matter is such that $w \geq 1$, the interesting cases to understand the relationship between energy density and anisotropic shear in the bounce regime are massless scalar field with $w = 1$, inflationary potential, which allows regime with $w \approx 1$, and ekpyrotic and ekpyrotic-like potentials which allow the ekpyrosis phase with $w \gg 1$.
 In every simulation, the initial conditions for gravitational phase space variables were given for $c_1, c_2, p_1, p_2, p_3$ while that of $c_3$ was determined by imposing the effective Hamiltonian constraint $\mathcal{H} = 0$ once the initial conditions for matter content were specified. These were given either by specifying the initial value of $\phi$ and $p_\phi$, or the value of $\phi$ at some specific energy density $\rho$. For the results presented below, 150 simulations were used for each type of matter content, and the majority of the simulations used randomized initial conditions from a uniform distribution within the given ranges:  $c_1, c_2 \in [-0.6,0.6]$, $\rho \in [0,0.01]$ and $\phi \in [0, 0.4]$. Additionally, the triad components were chosen from a uniform distribution of integers between 300 and 5000, and $c_3$ was solved for to ensure $\mathcal{H} = 0$. (All values above and in the following are in Planck units). These initial conditions were then used to obtain numerical solution using effective Hamilton's equations of motion for the triad and connection variables \eqref{triad_flow}.

\begin{figure}[tbh!]
    \centering
    \includegraphics[scale=0.75]{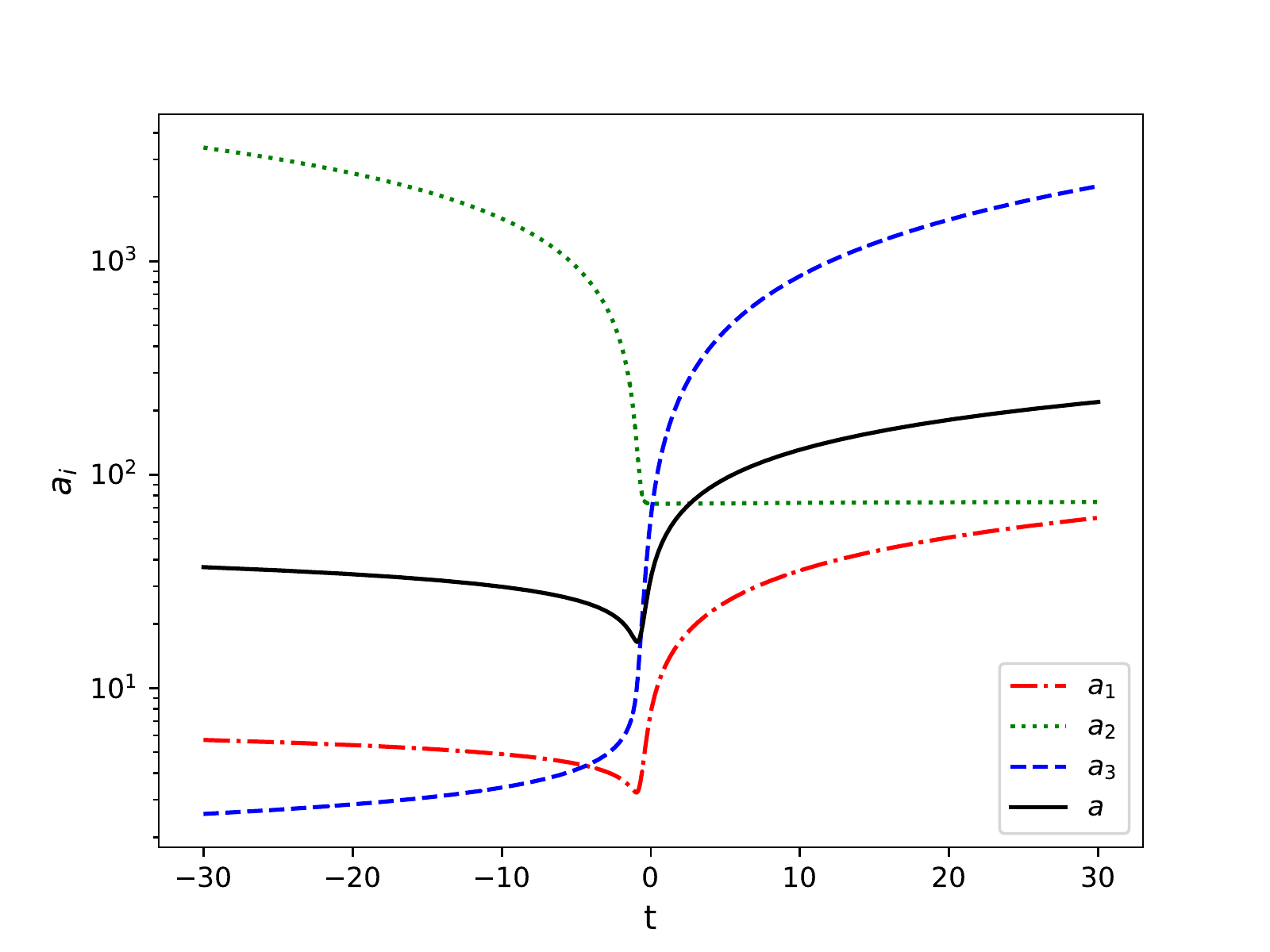}
    \caption{A typical evolution of the directional scale factors for a universe with a massless scalar field in Bianchi-I model of LQC is shown. All the scale factors undergo a non-singular bounce. The approach to bounce in the pre-bounce epoch is cigar like with two scale factors decreasing and one increasing. The same is true in the post-bounce epoch.
    The initial conditions are $c_1 = 0.57112, c_2 = 0.09828, p_1 = 4130, p_2 =630, p_3 = 2745, \phi=0.19234, \rho=0.00355$ and $c_3$ is determined from the vanishing of effective Hamiltonian constraint.}
    %solved to be $c_3 = 5.54912$}
    \label{fig:massless_scalar_scale_factors}
\end{figure}

\subsection{Massless Scalar Field}
\begin{figure}[t]%{1.0\textwidth}
\centering

\begin{subfigure}
    \centering
    \includegraphics[width=0.49\linewidth]{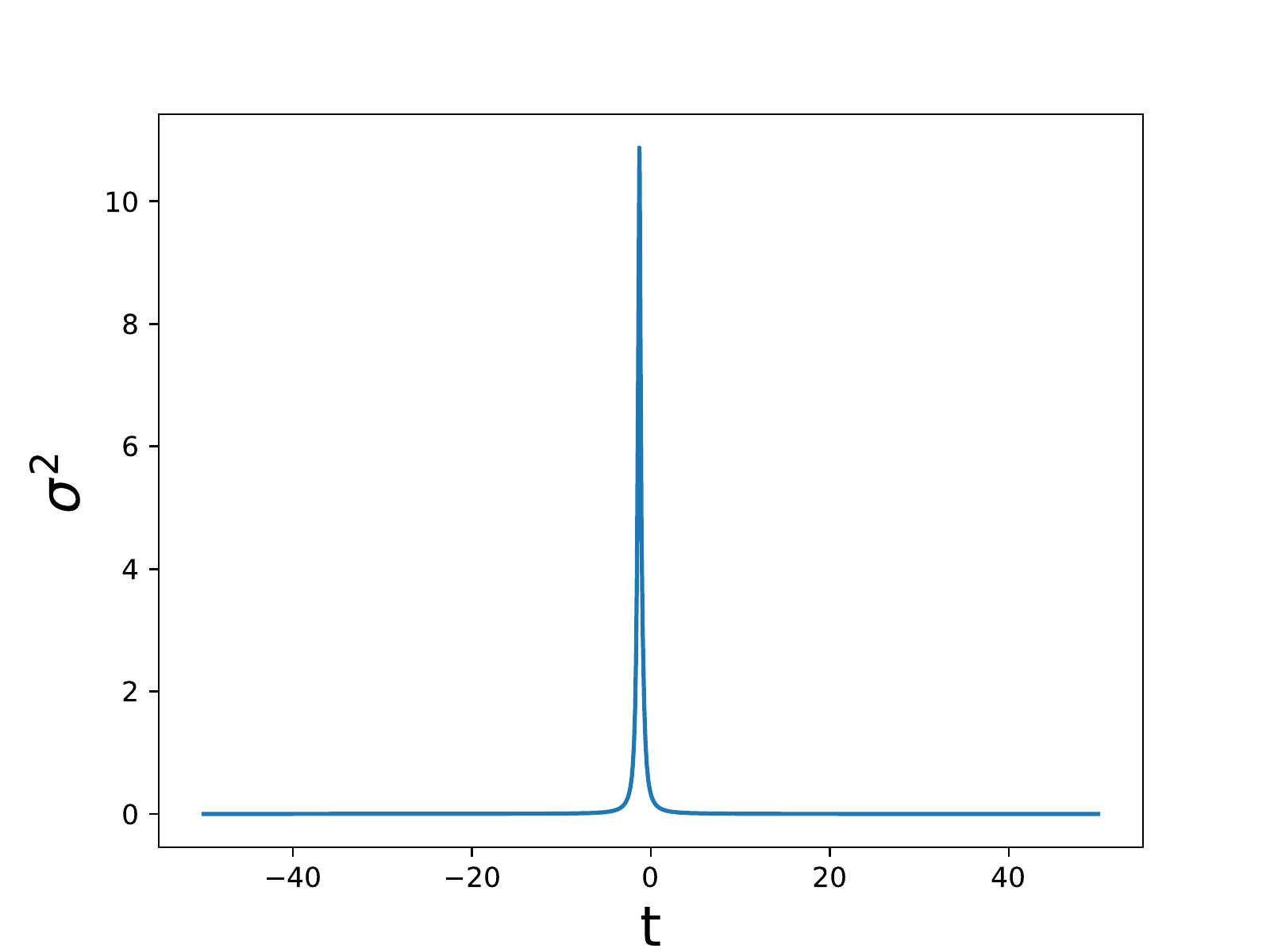}
    % \label{fig:massless_scalar_shear}
\end{subfigure}
\begin{subfigure}
    \centering
    \includegraphics[width=0.49\linewidth]{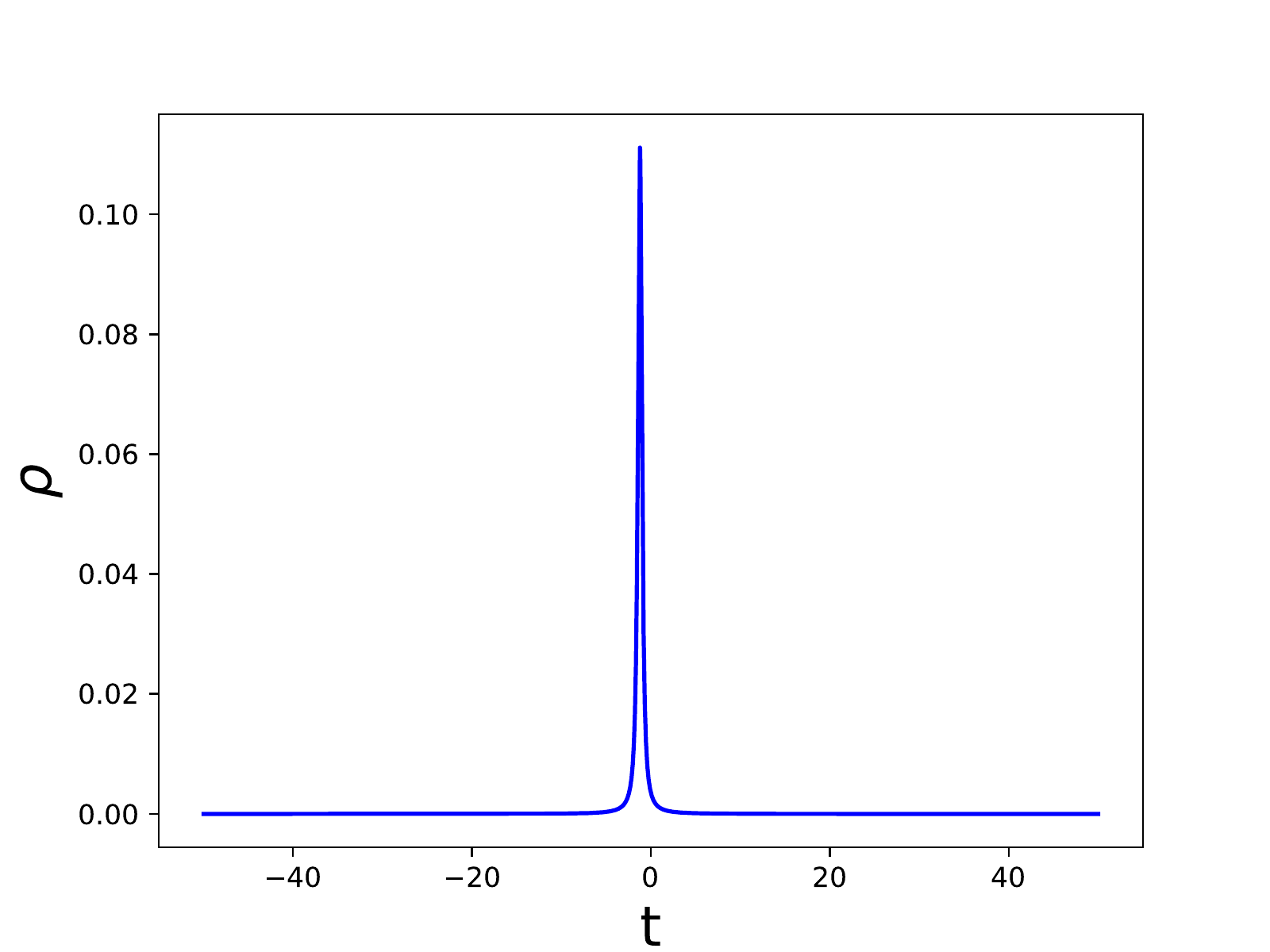}
    %\label{fig:massless_scalar_rho}
\end{subfigure}
\caption{Time evolution of anisotropic shear (left) and energy density (right) for the simulation corresponding to Fig. \ref{fig:massless_scalar_scale_factors}. Unlike classical theory, anisotropic shear and energy density are bounded in LQC.}
%The initial conditions are $c_1 = 0.57112, c_2 = 0.09828, p_1 = 4130, p_2 =630, p_3 = 2745, \phi=0.19234, \rho=0.00355$ and $c_3$ solved to be $c_3 = 5.54912$}
\label{fig:massless_scalar_shear_rho}
\end{figure}

We begin with the effective dynamics of the Bianchi-I spacetime with a massless scalar field as the matter content. The matter Hamiltonian in this case is
\begin{equation}
    \mathcal{H}_\mathrm{m} = \frac{p_{\phi}^2}{2v} =\frac{\dot{\phi}^2}{2}v,
\end{equation}
which using Hamilton's equations gives $p_{\phi} = \text{constant}$ and $P = \rho$, so that the equation of state is $w = 1$. The energy density $\rho = p_\phi^2/ 2v^2$ is proportional to $1/v^2$. Thus, $\rho$ has maxima precisely at the minima of $v$, i.e. at the bounce. We show a typical example of the numerical evolution in this case in Figs. \ref{fig:massless_scalar_scale_factors}  and \ref{fig:massless_scalar_shear_rho}. The evolution of the mean scale factor in effective dynamics in LQC, $a$ in Fig. \ref{fig:massless_scalar_scale_factors}, shows a non-singular bounce.  Fig. \ref{fig:massless_scalar_shear_rho} shows that energy density and the shear scalar both peak at the bounce. Note that the anisotropies dominate the bounce regime in this particular simulation, as the value of $\sigma^2/\sigma_{\mathrm{max}}^2$ is much greater than $\rho/ \rho_{\mathrm{max}}$ as seen in Fig. \ref{fig:massless_scalar_shear_rho}. While in this simulation $\sigma^2$ has a single peak at the bounce, this does not reflect a generic behavior since $\sigma^2$ is not monotonic in mean scale factor $a$. In fact, it is common for $\sigma^2$ to have an additional peak near the bounce.
Unlike the ekpyrotic and ekpyrotic-like cases considered in later subsections, we only find one bounce for the massless scalar field in our simulations. This is expected because, as the universe quickly becomes classical after the bounce, the classical Friedmann equation \eqref{Friedmann} is a good approximation in this regime and a turnaround in volume can only happen if $\rho=-\sigma^2 / 16\pi G$. This cannot happen in this case as both $\sigma^2$ and $\rho= p_\phi^2/ 2v^2$ are positive definite. As we will see for other matter content, this is not necessarily true for all the potentials we have considered.

\begin{figure}[h]
    \centering
    \includegraphics[scale=1.2]{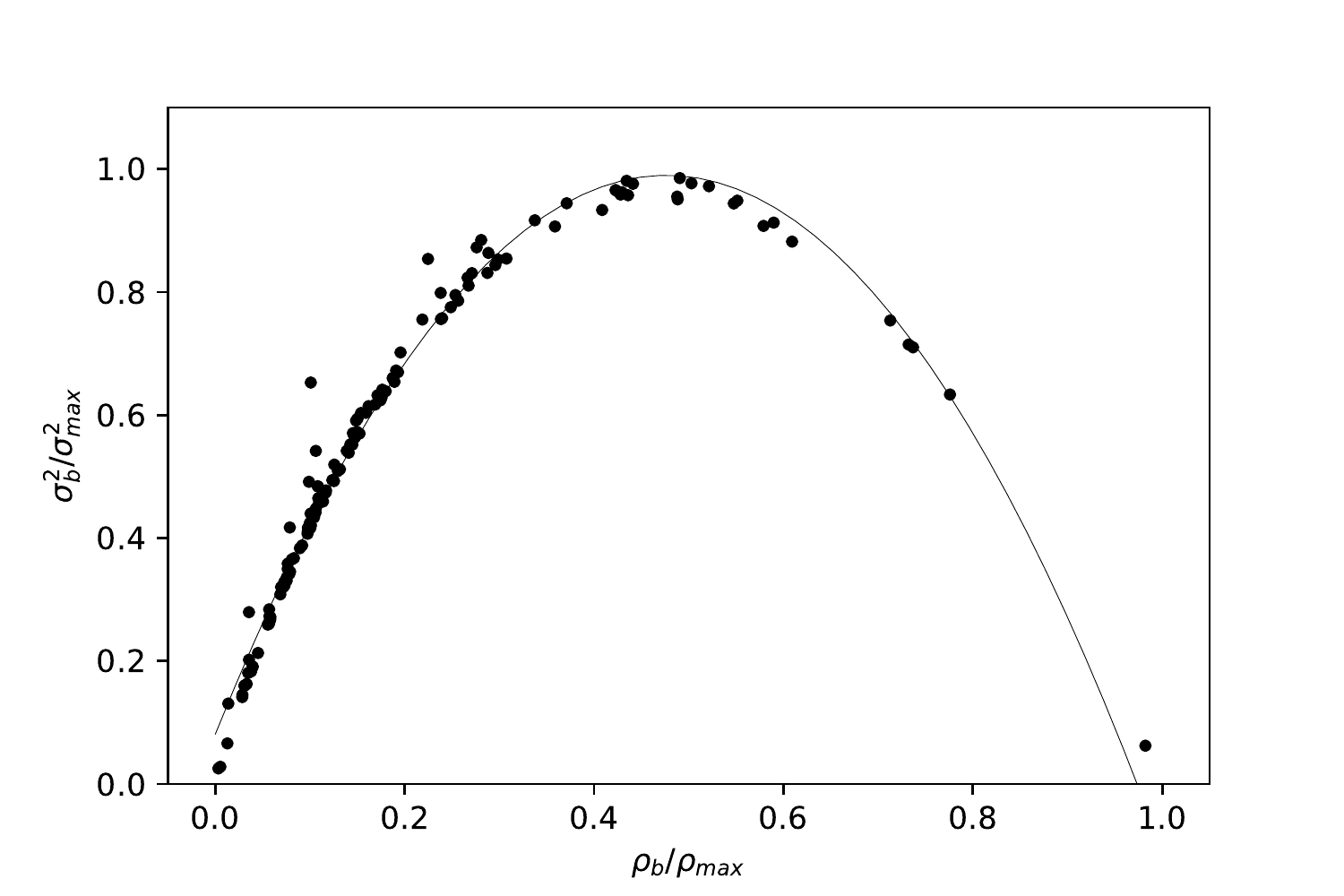}
    \caption{The ratio of anistropic shear at the bounce $(\sigma^2_b$) and the maximum allowed valued of anisotropic shear is plotted versus the ratio of energy density at the bounce $(\rho_b$) and its maximum allowed value for various numerical simulations with varying initial conditions. Each dot represents results from one simulation with randomized initial conditions.}
    \label{fig:massless_scalar_shear_vs_rho}
\end{figure}

So far we discussed only one example simulation of Bianchi-I model in LQC.
We now consider the relationship between $\rho$ and $\sigma^2$ at the bounce for Bianchi-I universe with a massless scalar field for a variety of initial conditions leading to more than 150 numerical simulations.  As discussed above, the anisotropic shear and energy density in LQC effective dynamics are bounded above. The results from our numerical simulations are plotted in Fig. \ref{fig:massless_scalar_shear_vs_rho}. The values obtained for $\sigma^2/\sigma_{\mathrm{max}}^2$ versus $\rho / \rho_{\mathrm{max}}$ at the bounce for various initial conditions for the case of massless scalar field are shown.

The resulting plot gives us crucial and previously unknown insights into the behavior of the effective dynamics of the Bianchi-I model in LQC in the vicinity of the bounce. The inverted parabolic relation obtained between $\sigma^2$ and $\rho$ at the bounce is surprisingly simple, considering  that a generalized effective Friedmann equation for Bianchi-I model in LQC is not known. We note that the ratio $\sigma_b^2/\sigma_{\mathrm{max}}^2$ has maximum value when the ratio $\rho_b/\rho_{\mathrm{max}}$ is nearly half. Later we will see that the same features are observed for different potentials.
%except under rather strong assumptions on the shear scalar being very small throughout the evolution as obtained in \cite{ChiouKV2007}.
Fitting the equation of a parabola, where as before $\rho_b$ denotes bounce density and $\sigma^2_b$ denotes anisotropic shear at the bounce,
\begin{equation}\label{parabola}
    \frac{\sigma_b^2}{\sigma_{\mathrm{max}}^2} = a\left(\frac{\rho_b}{\rho_{\mathrm{max}}}\right)^2 + b \left(\frac{\rho_b}{\rho_{\mathrm{max}}}\right) + c
\end{equation}
to the data in Fig. \ref{fig:massless_scalar_shear_vs_rho}, we find that $a \approx -4.0043, b \approx 3.8148$ and $c \approx 0.0812$.
%$a\approx-3.99$, $b\approx3.82$ and $c\approx 0.08$.
%$c\approx 0.0802$.
%Noting that there are larger deviations from the parabola at smaller values of $\rho$, it is likely that $c$ is even smaller. It appears that $a$ and $b$ are very similar in value, while $c$ is very close to zero.

Note that in contrast to these results from numerical simulations, the approximate expression for the generalized effective Friedmann equation derived in \cite{ChiouKV2007} under assumptions of low shear scalar gives a tilted-elliptic curve in the $\sigma^2$ versus $\rho$ plane at the bounce which does not match the data points obtained in Fig. \ref{fig:massless_scalar_shear_vs_rho}. This is because the assumption of low shear scalar  throughout the evolution can be severely limiting since the dynamics in Bianchi-I is more likely to be dominated by the shear scalar in the bounce regime as we have discussed above. In fact, for the set of initial conditions considered here, we find that the average values for the anisotropic shear and energy density at the bounce are $\sigma^2\approx6.1415$ and $\rho~\approx~0.0849$, which implies that on average $\sigma^2/\sigma_{\mathrm{max}}^2 > \rho/\rho_{\mathrm{max}}$ and the anisotropic shear often dominates over the energy density at the bounce when we consider matter content as a massless scalar field.

We further note that as a consequence of the parabolic relation, a larger energy density at the bounce is not always accompanied by a decrease in the anistropic shear at the bounce, and an opposite behavior can also occur. For example, if as a consequence of changing the initial conditions, the bounce point moves upwards toward the peak from the left half of the parabola, then an increase in both $\rho$ and $\sigma^2$ will be observed at the bounce. Another curious feature from Fig. \ref{fig:massless_scalar_shear_vs_rho} is the occurrence of bounces at relatively low values of both $\rho /\rho_{\mathrm{max}}$ and $\sigma^2 /\sigma_{\mathrm{max}}^2$, which lie on the left side of the parabola. We note that the values of energy density and shear scalar at these bounces are of the order of about $1\%$ of the Planckian values and are still very high compared to the classical limit. An important question is whether this novel parabolic relationship between the anisotropic shear and the energy density at the bounce is robust to change in the matter content. As we show in the following subsections, this parabolic behavior is obtained for all the matter fields considered in this manuscript, indicating that it may be a feature of the dynamics of the model itself.

\begin{figure}[h]
    \centering
    \includegraphics[scale=0.9]{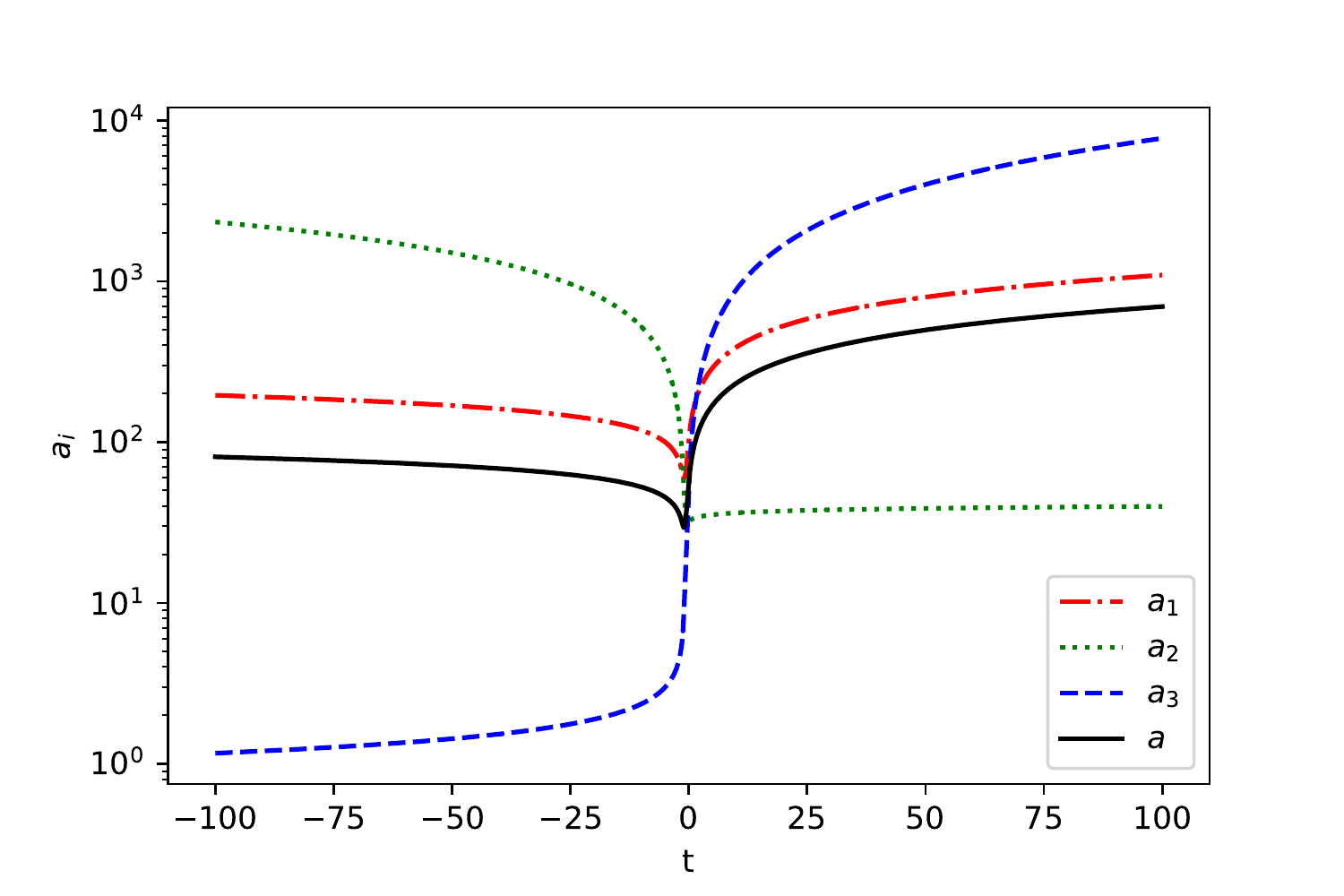}
    \caption{Evolution of scale factors for a universe with a massive scalar field in a $\phi^2$ potential is shown. The initial conditions are $c_1 = 0.0178, c_2 = 0.5631, p_1 = 1800, p_2 = 4431, p_3 = 2564, \phi=0.4569, P_\phi = -13124.6061$ and $c_3$ solved from Hamiltonian constraint. }
    \label{fig:inflation_scalar_scale_factors}
%actual Bianchi_c1=0.0178c2=0.56309c3=21.47074p1=1800p2=4431p3=2564phi=0.45686P_phi_0-13124.60613,   rho_o = 0.004211630737053006
\end{figure}

\subsection{Inflationary potential}
An important question in the inflationary paradigm is the way inflation onsets in presence of anisotropies. In particular, whether starting from generic initial conditions for anisotropic shear there exists an inflationary attractor. In LQC, this question has been answered earlier  where naturalness of inflation was explored in the Bianchi-I model in LQC \cite{GuptInflationB-I, barrau2}. For simplicity we consider the case of a  $\phi^2$ potential, and the results reported here are expected to be valid for other inflationary potentials too. It turns out that near the classical singularity the kinetic energy term of the scalar field generally becomes dominant compared to the potential energy. For this reason, results from the inflationary potential turn out to be quite similar to the case of the massless scalar field.
%This is an important case as its often used in inflationary models of the early universe. Since the initial state of the universe in general could be anisotropic, it is important to consider this potential in the Bianchi-I model.

\begin{figure}[h]
    \centering
    \includegraphics[scale=1.15]{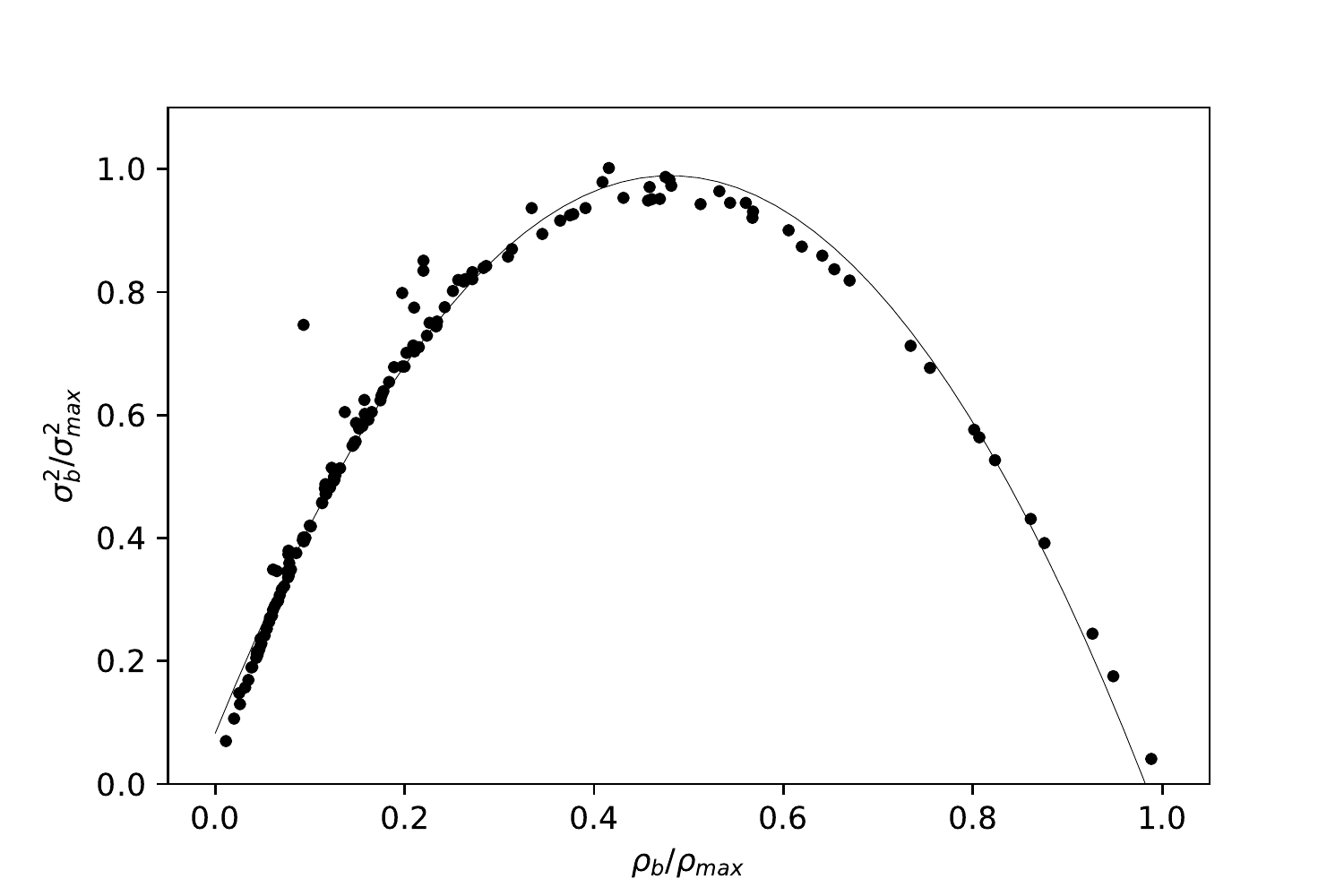}
    \caption{Anisotropic shear vs energy density at at the bounce for 150 numerical simulations with varying initial conditions for $m^2 \phi^2$ potential with $m = 0.001$. }
    \label{fig:inflation_scalar_shear_vs_rho}
\end{figure}

The Hamiltonian for the scalar field in this case is given by
\begin{equation}
    \mathcal{H}_\mathrm{m} = \frac{p_{\phi}^2}{2v} + \frac{1}{2}m^2\phi^2.
\end{equation}
Since we are interested in finding qualitative behavior of bounce we choose for simplicity of numerics $m = 0.001$.
Using effective dynamical equations a typical evolution of the directional scale factors and the mean scale factor is shown in Fig. \ref{fig:inflation_scalar_scale_factors}. The bounce being a generic feature of the dynamics of the effective Bianchi-I spacetime, is also found in this case with a similar behavior for energy density and anisotropic shear as in the massless case.  Similarly to the massless case, the energy density in this case is also positive definite. Thus, there is no turning point for dynamics in the classical regime as per the classical generalized Friedmann equation \eqref{Friedmann}, so we see only a single bounce in this case as well. The plots for the evolution of the shear scalar and the energy density in this case are similar to those of the massless case shown in  Fig. \ref{fig:massless_scalar_shear_rho}.

%actual Bianchi_c1=0.0178c2=0.56309c3=21.47074p1=1800p2=4431p3=2564phi=0.45686P_phi_0-13124.60613

We now discuss the relationship of $\rho$ and $\sigma^2$ at the bounce in the inflationary case. For this, we perform a large number of numerical simulations with varying initial conditions as described above. The data for the energy density and the shear scalar at the bounce in this case is plotted in Fig. \ref{fig:inflation_scalar_shear_vs_rho} for 150 simulations. As in the case of the massless scalar field, the relationship between $\sigma^2_b /\sigma^2_{\mathrm{max}}$ and $\rho_b /\rho_{\mathrm{max}}$  turns out to be parabolic with the maximum value of $\sigma_b^2/\sigma_{\mathrm{max}}^2$ occurring, as in the massless scalar case, at $\rho_b/\rho_{\mathrm{max}} \approx 1/2$. Fitting the parabolic equation \eqref{parabola} to the data in Fig. \ref{fig:inflation_scalar_shear_vs_rho} yields the values $a \approx -3.9279, b \approx 3.7739$ and $c \approx 0.0828$ for the parameters of the parabola, which are very close to those observed for the massless case in Fig. \ref{fig:massless_scalar_shear_vs_rho}. As in the case of the massless scalar field, the distribution of  points tells us that the bounces in general are not occurring at values of energy density close to its maximum value and hence anisotropic shear plays an important role in singularity resolution.

% actual [-3.92791691  3.77387187  0.08276501]

\subsection{Ekpyrotic potential}
In this and the next subsection, we consider two slightly different potentials, named ekpyrotic and ekpyrotic-like potentials which are considered in models for alternatives of inflation. The ekpyrotic/cyclic model is motivated by M-theory where the inter-brane separation is determined by a moduli field, $\phi$, with an ekpyrotic potential given as \cite{turok-steinhardt}

\begin{equation}\label{Ekpyrotic_potential}
    U_1(\phi) = U_{1_o}\left(1-e^{-\sigma_1\phi}\right)\exp\left({-e^{-\sigma_2\phi}}\right),
\end{equation}
%\vskip-0.5cm
where $U_{1_o}$, $\sigma_1$ and $\sigma_2$ are parameters of the potential. In this model, $\phi$ slowly rolls in an almost flat and positive portion of the potential as the branes move away from each other, but becomes steep and negative as the branes begin approaching each other and the visible universe begins to contract. While the model was originally motivated from brane dynamics in a bulk spacetime, the above potential is generally used as an effective potential in cosmological models.
The shape of the potential is such that it can lead to a phase of ekpyrosis in which the equation of state is $w = P/\rho \gg 1$.  This can result in  isotropization near the classical singularity. % \cite{turok-steinhardt1, Buchbinder:2007ad}. %Similar features are demonstrated by an ekpyrotic-like potential \cite{CaiBranden2012} discussed in the next subsection.
Previous work in LQC using this potential shows that singularity is resolved in isotropic \cite{Bojowald:2004kt,Singh:2006im} as well as Bianchi-I spacetime \cite{CalitauPSKV2009}. However, details of the interplay between the energy density and anisotropic shear in the bounce regime have so far remained unexplored.

\begin{figure}[H]
    \centering
    \includegraphics[scale=0.7]{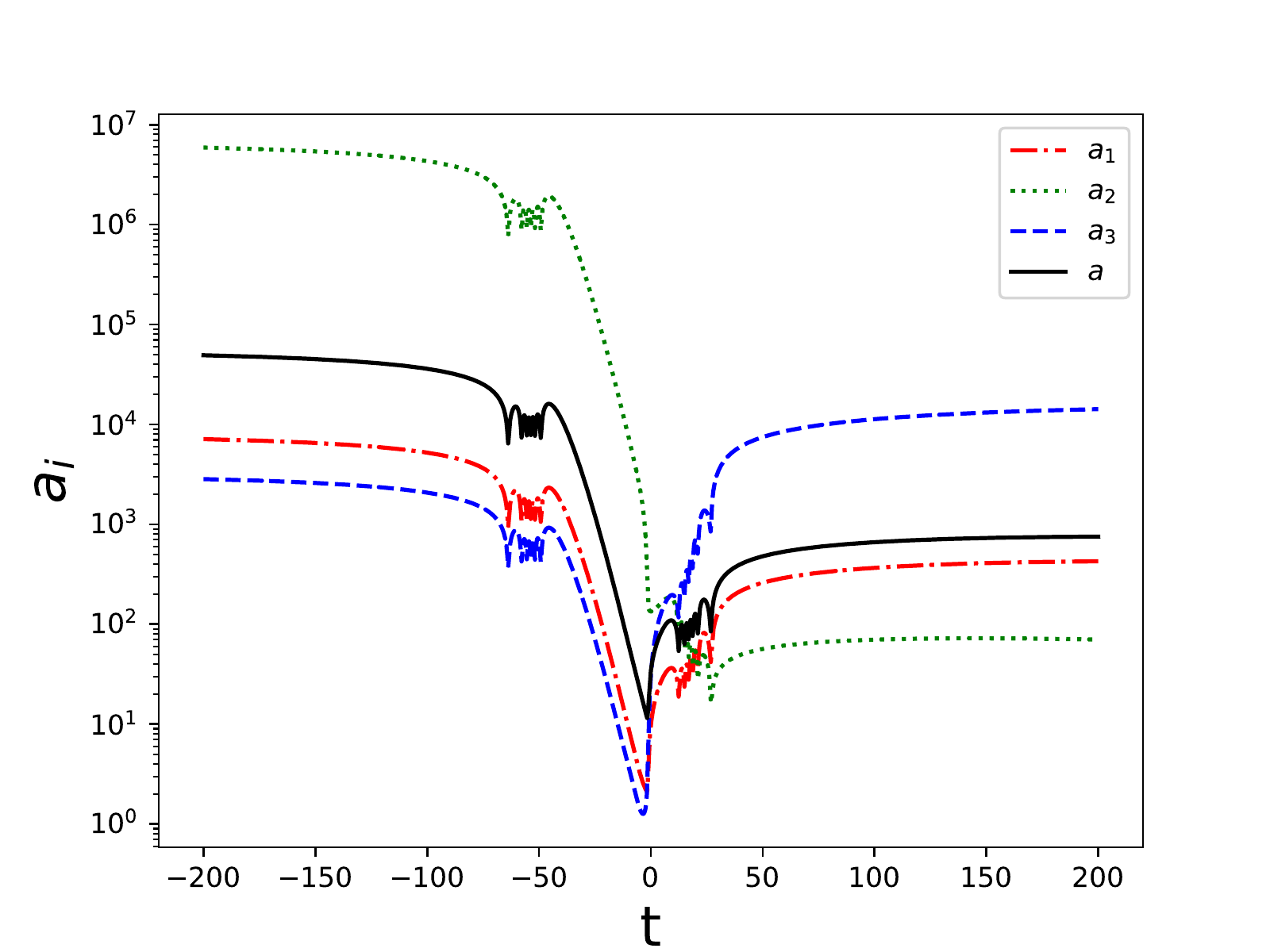}
    \caption{Behavior of scale factors for the ekpyrotic potential \eqref{Ekpyrotic_potential} in a typical simulation is shown. The initial conditions are $c_1 = 0.57112, c_2 = 0.09828, p_1 = 4130, p_2 =630, p_3 = 2745, \phi=0.19234, \rho=0.00355$ and $c_3$ is determined from the vanishing of Hamiltonian constraint at the initial surface.}
    \label{fig:ogEkpyrotic_scale_factors}
\end{figure}

We now consider numerical simulations of Bianchi-I spacetimes with the ekpyrotic potential given in \eqref{Ekpyrotic_potential} in this subsection. The matter component of the Hamiltonian in this case is given by,
\begin{equation}
    \mathcal{H}_m = \frac{p_{\phi}^2}{2v} + U_1(\phi)v.
\end{equation}

\begin{figure}[t!]
\centering
\begin{subfigure}
    \centering
    \includegraphics[width=0.49\linewidth]{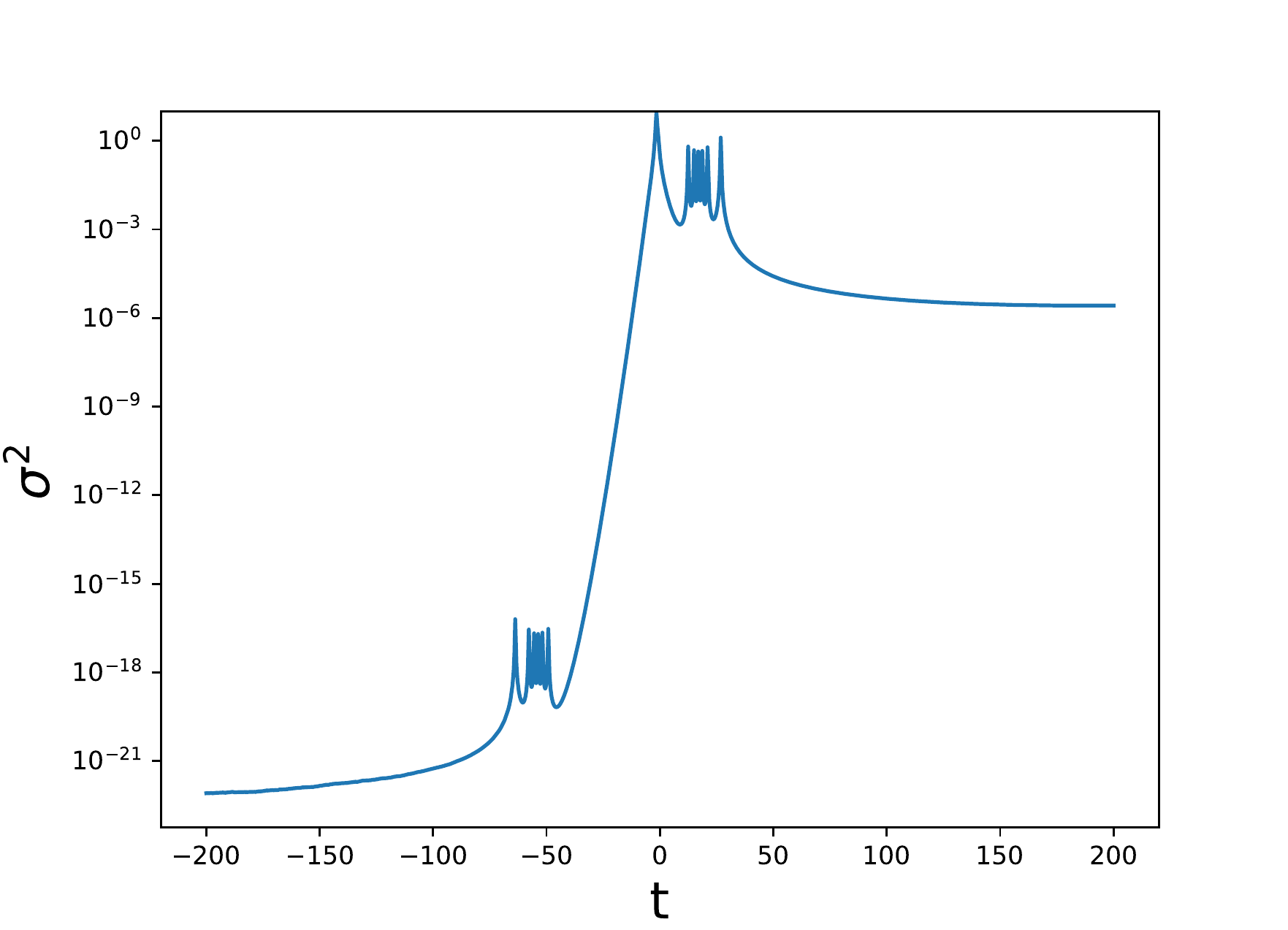}
    % \caption{Evolution of anistropic shear for Ekpyrotic potential and initial conditions $c_1=0.587, c_2=0.144, p_1 = 4896, p_2 = 377, p_3=4371, \phi=0.24387, \rho=$ and $c_3$ solved to be $c_3=2.559$}
    \label{fig:ogEkpyrotic_shear}
\end{subfigure}
\begin{subfigure}
    \centering
    \includegraphics[width=0.49\linewidth]{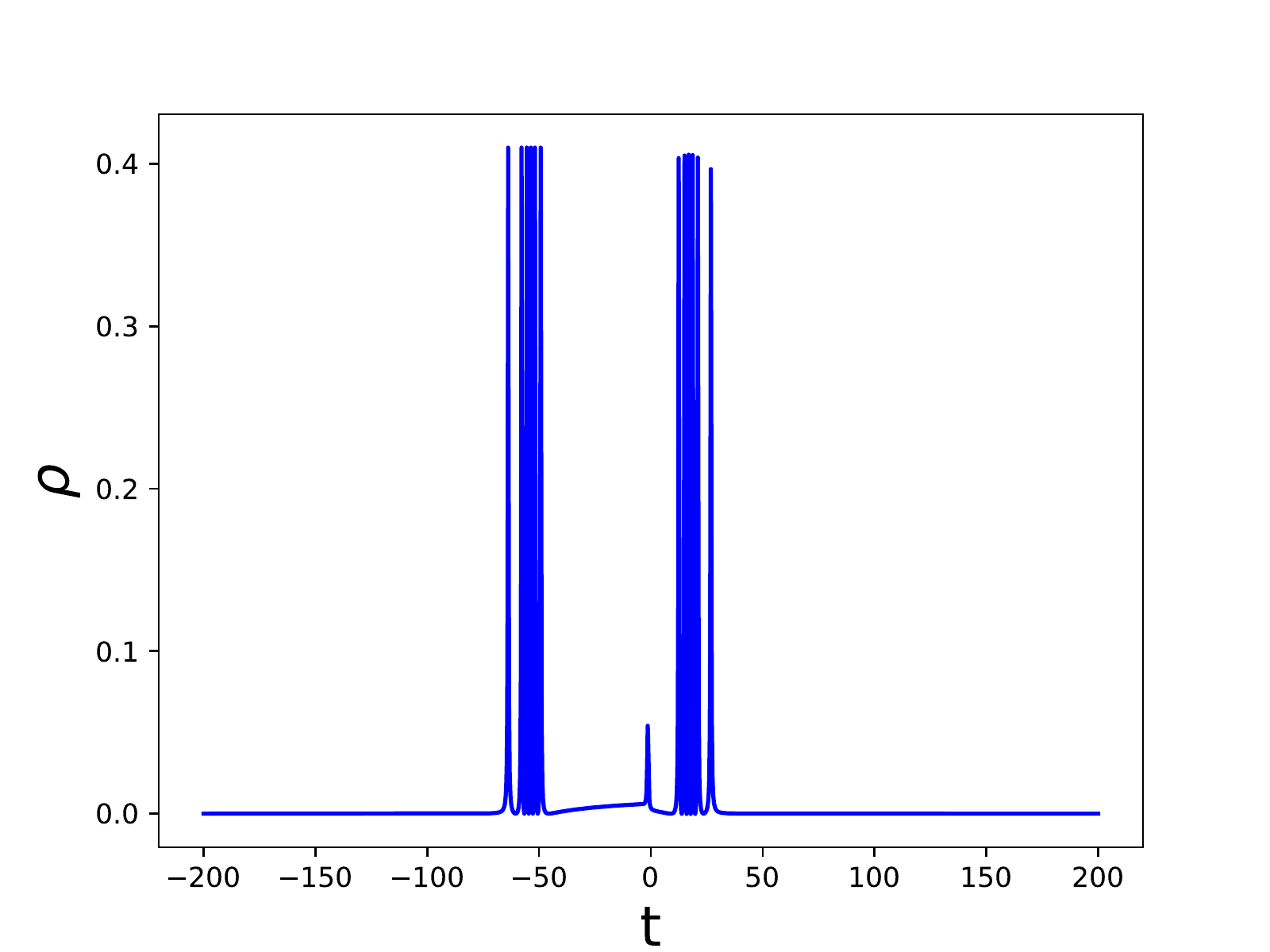}
    % \caption{Evolution of energy density for Ekpyrotic potential and initial conditions $c_1=0.587, c_2=0.144, p_1 = 4896, p_2 = 377, p_3=4371, \phi=0.24387, \rho=$ and $c_3$ solved to be $c_3=2.559$}
    \label{fig:ogEkpyrotic_rho}
\end{subfigure}
\caption{Evolution of anistropic shear (left) and energy density (right) for the ekpyrotic potential. Results correspond to the simulation shown in Fig. \ref{fig:ogEkpyrotic_scale_factors}.}
\label{fig:ogEkpyrotic_shear_rho}
\end{figure}

Simulations of the dynamics in this case are performed for a variety of initial conditions with the potential parameters taken to be $U_{1_o}=0.02,$ $\sigma_1 = 0.3\sqrt{8\pi}$ and $\sigma_2 = 0.09\sqrt{8\pi}$. We note that similar results have been obtained for variations in the value of these parameters. The behavior of the scale factors, anisotropic shear and energy density in this case are shown in Fig. \ref{fig:ogEkpyrotic_scale_factors}-\ref{fig:ogEkpyrotic_shear_rho}.
%for initial conditions $c_1=0.57112, c_2=0.09828, p_1=4130, p_2=630, p_3=2745, \phi=0.19234, \rho=0.00355$ and $c_3$ solved to satisfy the Hamiltonian constraint at the initial surface.

In contrast to the previous two cases of massless scalar field and inflationary potential, each individual scale factor, as well as the mean scale factor $a=\left(a_1a_2a_3\right)^{1/3}$ can bounce many times as can be seen in Fig. \ref{fig:ogEkpyrotic_scale_factors}. This is because the energy density, $\rho$  in this case is not positive definite and a recollapse occurs in the classical regime whenever the negative value of the energy density balances the contribution of the shear scalar in the classical generalized Friedmann equation \eqref{Friedmann}. We note that due to the effect of the ekpyrotic potential, the universe seems to have isotropized after the last bounce as the scale factors are seen to be evolving with almost the same rate after the last bounce. Corresponding to the multiple bounces seen in the scale factors, there are a large number of peaks in the anisotropic shear scalar and energy density plots. The peaks in the energy density and shear scalar do not necessarily take place exactly at the bounce, but generally happen in the vicinity of the bounce. In Fig. \ref{fig:ogEkpyrotic_shear_rho} we see that most of the peaks in the anisotropic shear are extremely small with $\sigma^2 \ll \sigma_{\mathrm{max}}^2$, and we also find that most of the peaks have $\rho$ very close to $\rho_{\mathrm{max}}$. This implies that most bounces are highly isotropic, apart from the larger middle bounce which corresponds with the highest $\sigma^2$ peak and lowest $\rho$ peak.

% \begin{figure}[H]
%     \centering
%     \begin{subfigure}%{0.5\textwidth}
%     \centering
%     \includegraphics[width=0.4\linewidth]{Figures/ogEkpyrotic shear.pdf}
%     % \label{fig:Ekpyrotic_shear}
%     \end{subfigure}
%     \begin{subfigure}%{0.5\textwidth}
%     \centering
%     \includegraphics[width=0.4\linewidth]{Figures/ogEkpyrotic energy density.pdf}
%     % \label{fig:Ekpyrotic_energy_density}
%     \end{subfigure}
%     \caption{Evolution of shear scalar and Energy density near bounces}
%     \label{bounded_energy_shear_Ekpyrotic}
% \end{figure}

%\begin{figure}[H]
%    \centering
    % \includegraphics[scale=0.75]{Figures/ogEkpyrotic shear vs rho colored by outliers.pdf}
%    \includegraphics[scale=0.75]{Figures/Ekpyrotic_V0=0.02_parabola.png}
%    \caption{Shear scalar vs energy density at the last bounce with an Ekpyrotic potential. Points are marked according to whether $\rho$ and $\sigma^2$ individually increased or decreased with the potential as compared to the massless scalar case.}
%    \label{fig:ogEkpyrotic_shear_vs_rho}
%\end{figure}

\begin{figure}[H]
    \centering
    \includegraphics[scale=0.7]{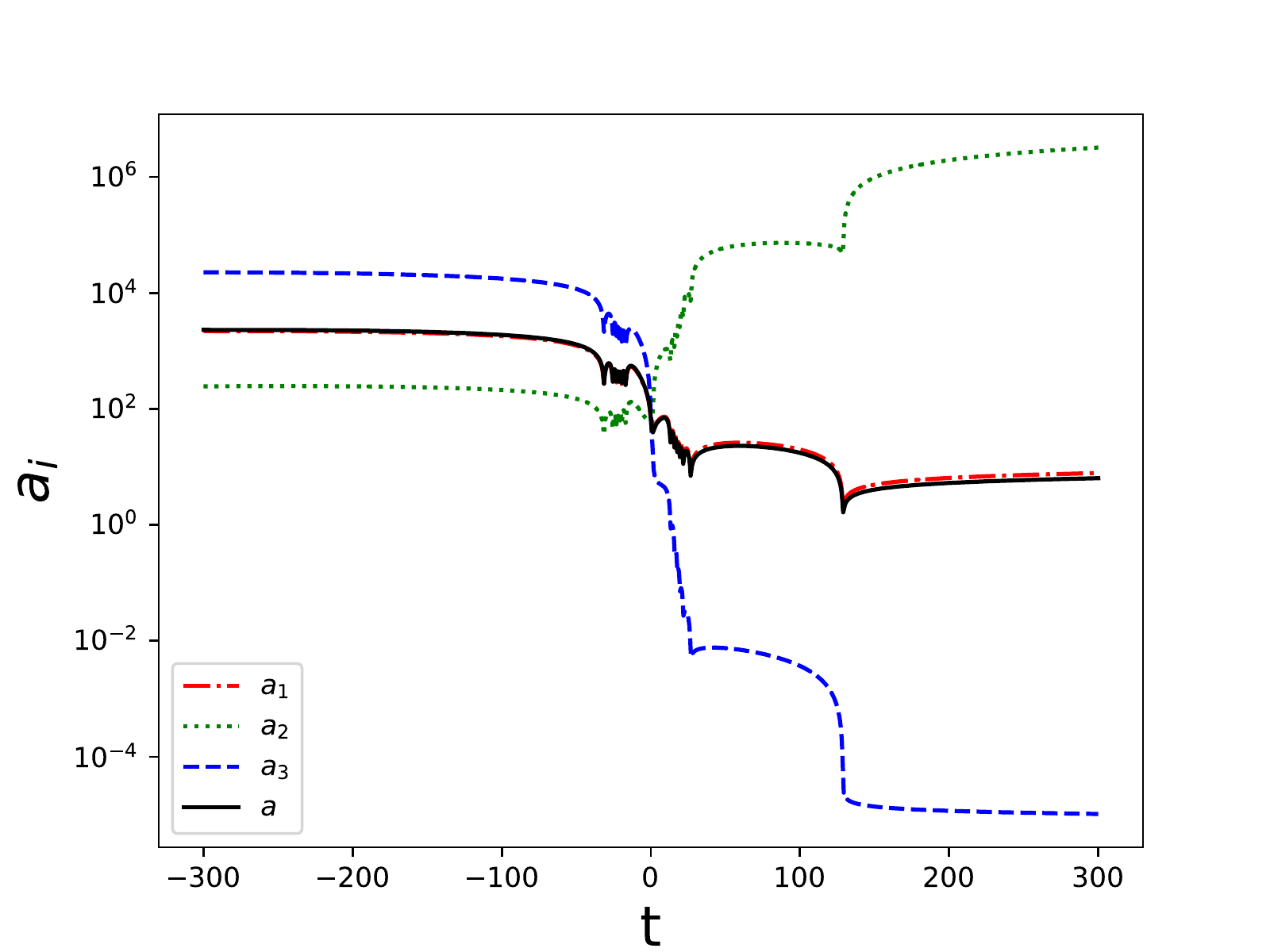}
    \caption{Behavior of scale factors for the ekpyrotic potential \eqref{Ekpyrotic_potential} in a typical simulation is shown. The initial conditions are $c_1 = -0.0201, c_2 = -0.38704, p_1 = 4413, p_2 =4309, p_3 = 3194, \phi=0.17514, \rho=0.00215$ and $c_3$ is determined from the vanishing of Hamiltonian constraint at the initial surface. Note that approach to bounces is point like as well as cigar like such as the prominent bounce neat $t \approx 0$ where $a_1$ and $a_3$ have opposite behavior than $a_2$. } \label{fig:cigar-ekpyo}
\end{figure}
Note that in the presence of ekpyrosis, since the equation of state $w$ can be larger than unity, the energy density of the scalar field can grow faster than $a^{-6}$ during the contracting phase. Energy density can thus dominate over the anisotropic shear in the dynamics and the approach to the bounce becomes point-like. We see this behavior in Fig. \ref{fig:ogEkpyrotic_scale_factors}. However, this does not imply that in the ekpyrotic case the approach to singularity in the classical theory and the bounce in LQC can not be cigar-like. The shape of the singularity and the resulting bounce is determined by the initial conditions and the parameters in the potential. As an example, in Fig. \ref{fig:cigar-ekpyo} we see that the approach to some bounces is point-like as well as cigar-like.

\begin{figure}[H]
    \centering
 \includegraphics[scale=1.1]{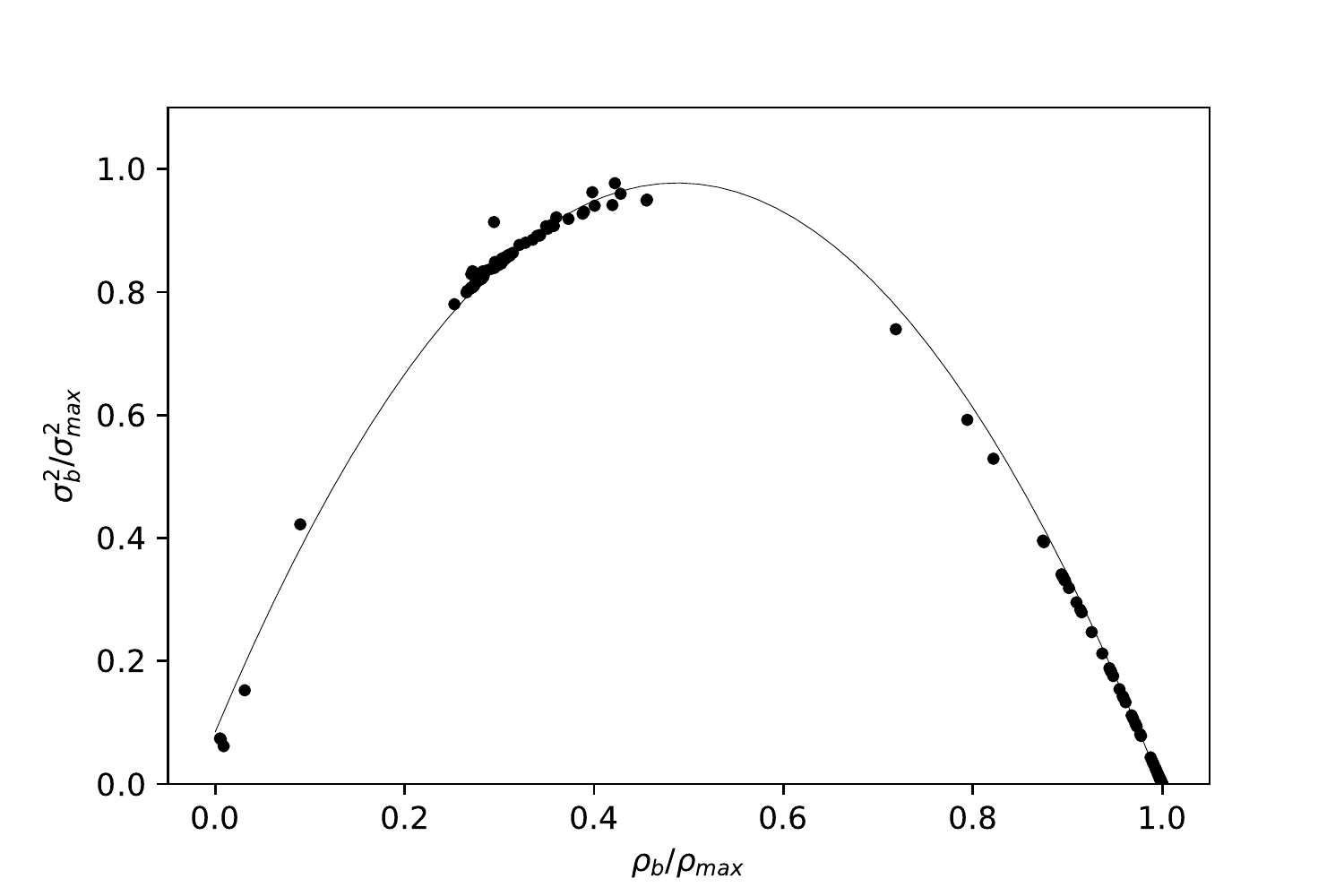}
    \caption{Shear scalar vs energy density at the last bounce in the time range for an ekpyrotic potential. Results are similar if some other bounce out of the multiple bounces is chosen.}
    \label{fig:ogEkpyrotic_shear_vs_rho}
\end{figure}

We performed many simulations for different initial conditions for the ekpyrotic potential and results from 150 such simulations are shown in Fig. \ref{fig:ogEkpyrotic_shear_vs_rho} where we plot
the value of shear scalar versus energy density at the last bounce in the given time range of evolution, generally taken as $t= [-500 , 500]$ in Planck units. Unlike the massless scalar case there are now multiple bounces and choice has to be made about which bounce to consider. While one could consider the first or the most prominent bounce, the last bounce is the final non-trivial quantum gravity event  before the expanding phase starts and for this reason is more interesting to understand isotropization of post-bounce branch. It should be noted that the results shown in Fig. \ref{fig:ogEkpyrotic_shear_vs_rho} (and similarly in next subsection) were qualitatively similar if instead we chose the first bounce, the most prominent bounce or some intermediate bounce.
As in the case of the massless scalar field and inflationary potential, we find that the points lie on a parabola. Fitting a parabolic equation of the form \eqref{parabola} in this case gives $a\approx-3.7494$, $b\approx3.6582$ and $c\approx 0.0851$, which matches closely with the fits found in the massless scalar field and inflationary case. As in the previous two cases, we find that $\sigma_b^2/\sigma_{\mathrm{max}}^2$ takes its maximum value when the ratio  $\rho_b/\rho_{\mathrm{max}} \approx 1/2$. Changing the parameters of the potential has only a little affect on the values of the fit parameters. We see that the parabolic relation between energy density and shear scalar at the bounce seems to survive even when we have matter with equation of state higher than unity, where the bounce is more likely to be dominated by energy density as compared to the shear scalar, in contrast to the previous cases. As a result, in this case, we also see an isotropization effect of the ekpyrotic potential as many points in Fig. \ref{fig:ogEkpyrotic_shear_vs_rho} lie on the right half of the parabola indicating higher energy density at the bounce. However, due to the parabolic relation, a higher energy density at the bounce may be accompanied by an increased shear scalar at the bounce in some cases. These correspond to the bounce points lying near the peak of the parabola which have both energy density and the shear scalar comparable to their maximum values.

% \begin{figure}[h]
%     \centering
%     \includegraphics[scale=0.75]{Figures/ogEkpyrotic shear over rho progression.pdf}
%     \caption{Progression of $\left|\sigma^2/\rho\right|$ from one bounce to the next, normalized so that $\left|\sigma^2/\rho\right|$ has a maximum at 1}
%     \label{fig:ogEkpyrotic_shear_over_rho_progression}
% \end{figure}

\begin{figure}[H]
    \centering
    \includegraphics[scale=0.8]{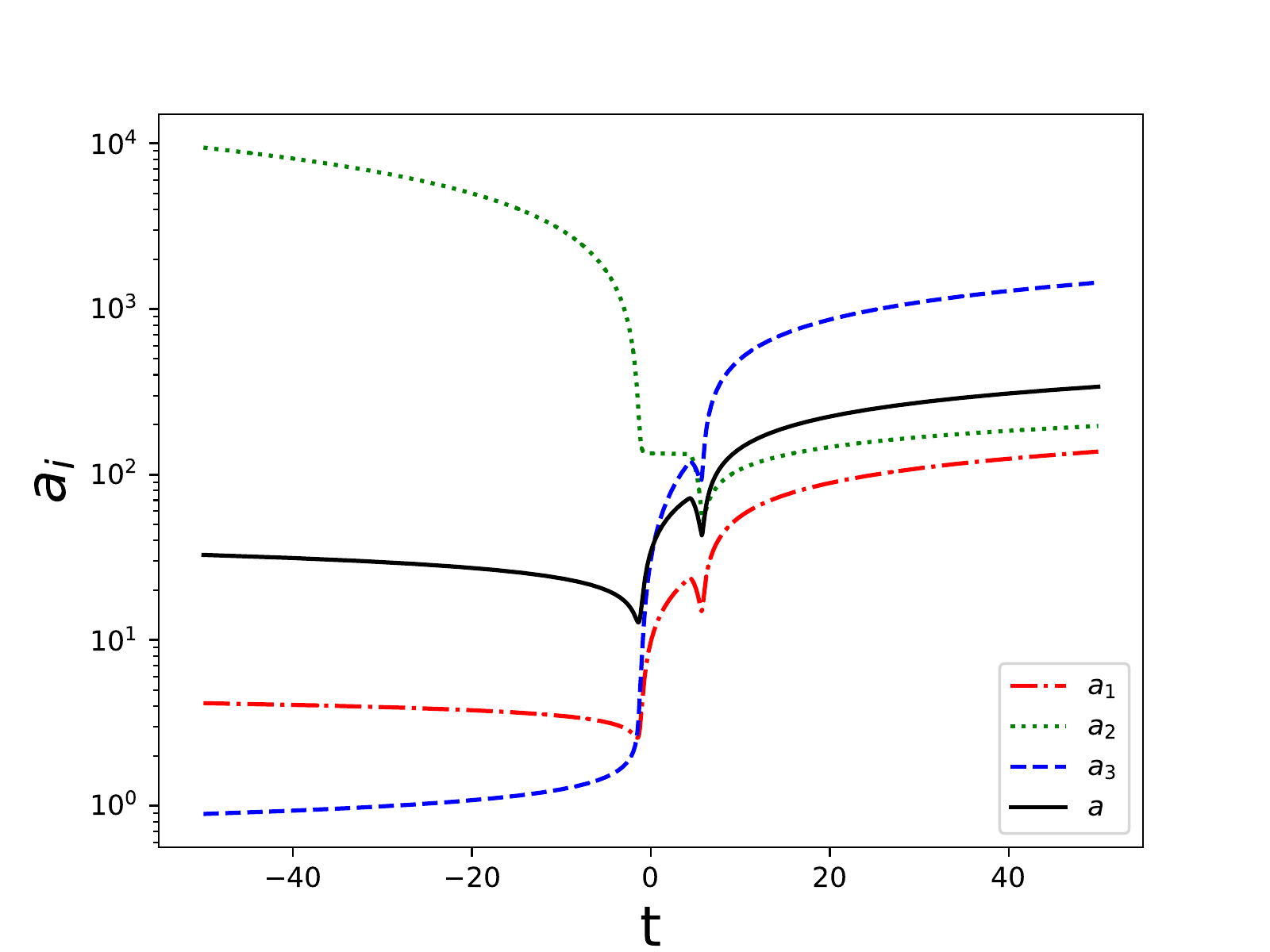}
    \caption{Behavior of scale factors from a  simulation in an ekpyrotic-like potential \eqref{potential}.  The initial conditions are $c_1 = 0.57112, c_2 = 0.09828, p_1 = 4130, p_2 =630, p_3 = 2745, \phi=0.19234, \rho=0.00355$ and $c_3$ determined from the vanishing of effective Hamiltonian constraint.}
    \label{fig:Ekpyrotic_like_scale_factors}
\end{figure}
\subsection{Ekpyrotic-like potential}\label{Ekpyrotic_like}
We now consider the effective Bianchi-I spacetime with an ekpyrotic-like potential used in \cite{CaiBranden2012}. This potential also shows the isotropization effects seen for the ekpyrotic potential in a bouncing universe. The phenomenological consequences of this potential for the background evolution and the power spectrum of curvature perturbations have been recently explored in detail in \cite{LiSainiSingh2021}. The potential is given by,
%\hskip-2cm
%\vskip-1.5cm
\begin{equation}\label{potential}
U_2(\phi)=-\frac{2 \, U_{2_o} }{e^{-\sqrt{\frac{16\pi}{p}}\phi}+e^{\beta\sqrt{\frac{16\pi}{p}}\phi}},
\end{equation}
%\vskip-1cm
with the parameters $U_{2_o}$, $p$ and $\beta$  all taking positive values. The matter Hamiltonian in this case becomes
%\vskip-2cm
\begin{equation}
    \mathcal{H}_\mathrm{m} = \frac{p_{\phi}^2}{2v} + U_2(\phi)v,
\end{equation}
%\vskip-0.1cm
where we take the potential parameters to be $U_{2_o} = 0.0366$, $p=0.1$ and $\beta=5$ in the simulations presented here, with similar results obtained for variations in values of these parameters.
%The energy density in this case, $\rho = \frac{p_\phi^2}{2v^2} + U_2(\phi)$, does not necessarily a maximum precisely at the bounce. However, both $\rho$ and $\sigma^2$ obtain maxima near the bounce.
%These parameters are chosen so that the minimum value of the potential matches that of the minimum potential value in the ekpyrotic-like potential from the previous section.

\begin{figure}[tbh!]
\centering
\begin{subfigure}
    \centering
\includegraphics[width=0.49\linewidth]{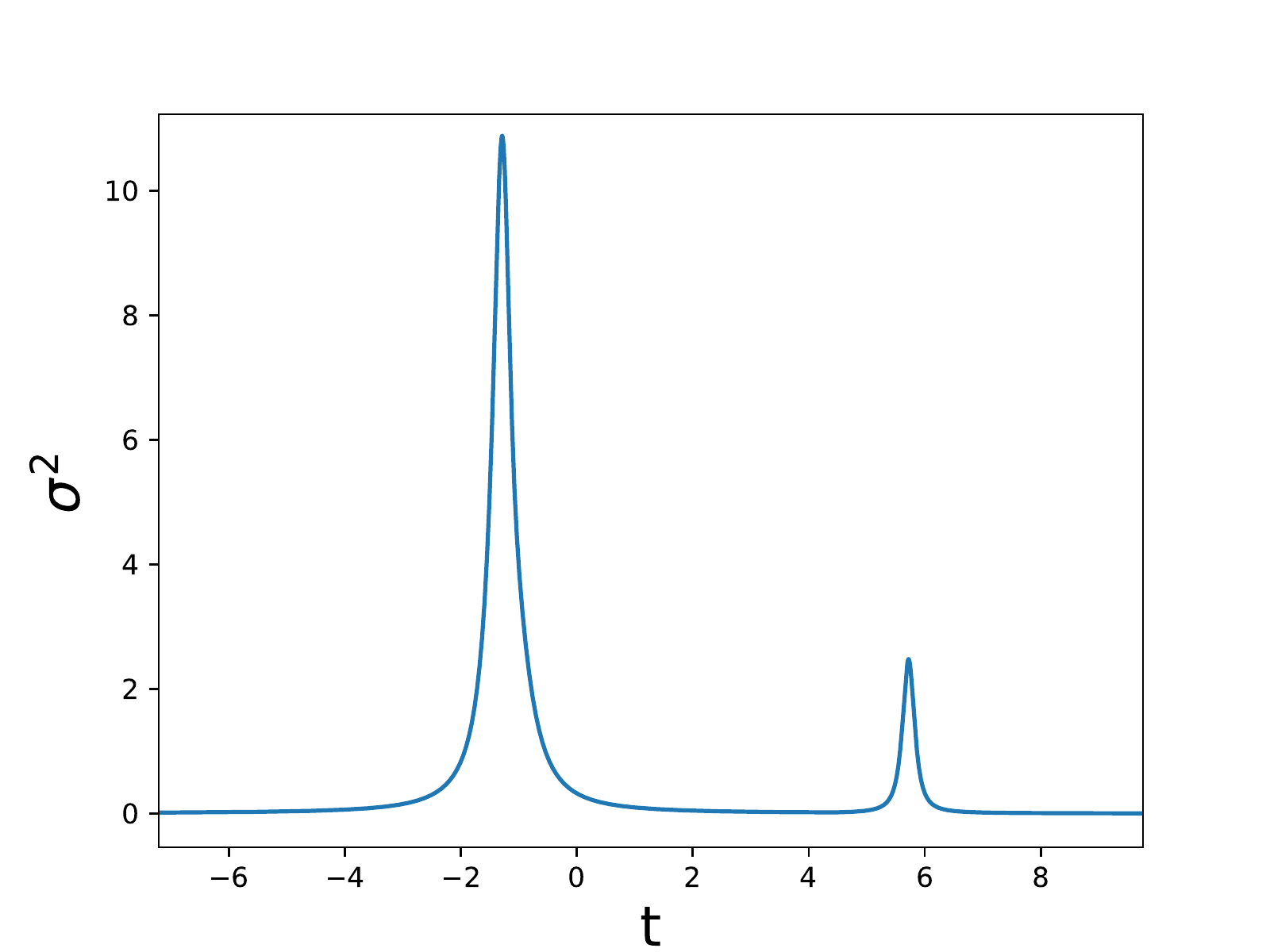}
    % \caption{Evolution of anistropic shear in ekpyrotic-like potential with initial conditions $c_1=-0.6, c_2=-0.5, p_1=500,p_2=360, p_3=340,\phi=0.4, \rho=$ and $c_3$ solved to be $c_3=0.3094$}
    \label{fig:Ekpyrotic_like_shear}
\end{subfigure}
\begin{subfigure}
    \centering
    \includegraphics[width=0.49\linewidth]{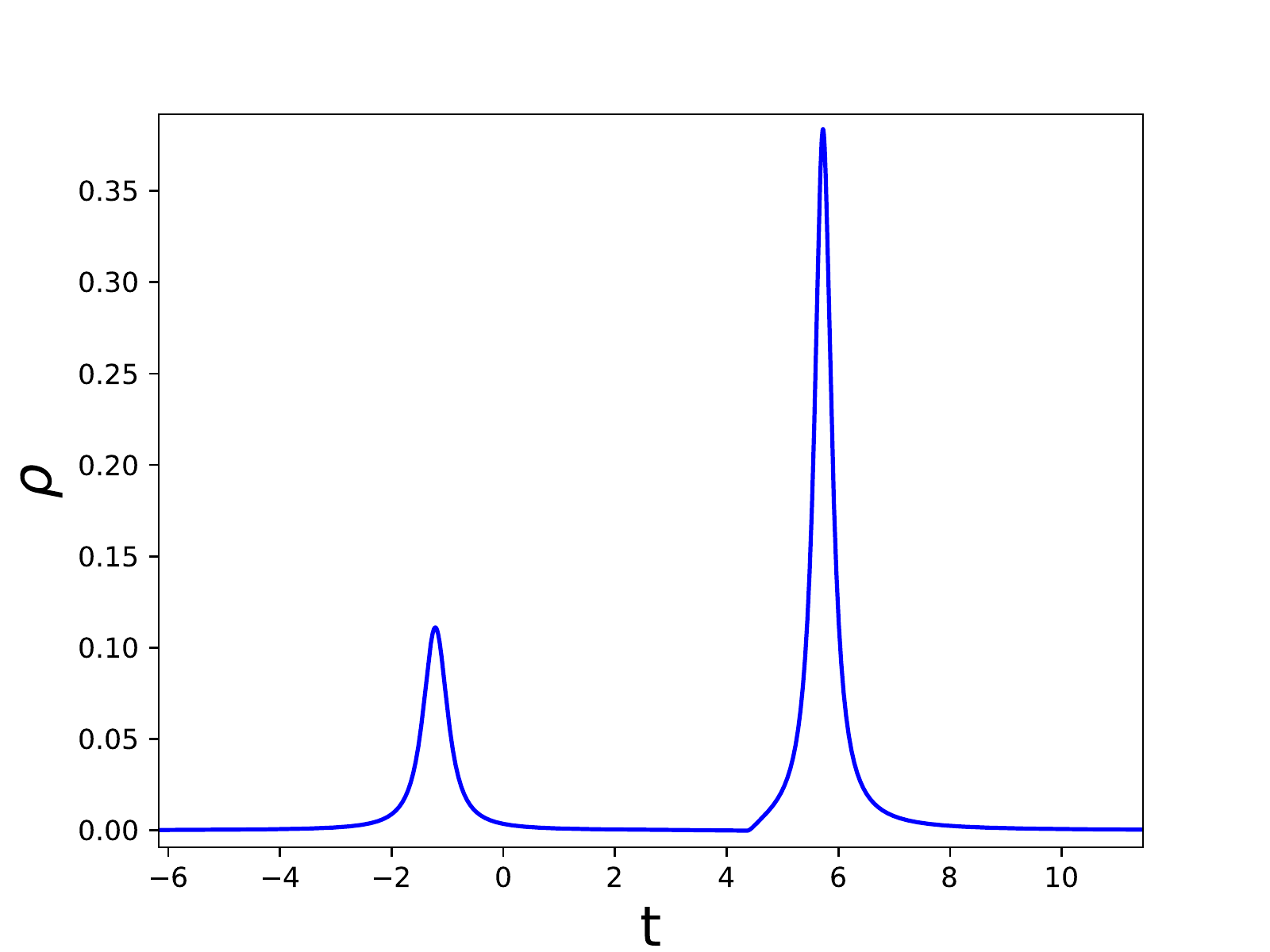}
    % \caption{Evolution of energy density in ekpyrotic-like potential with initial conditions $c_1=-0.6, c_2=-0.5, p_1=500,p_2=360, p_3=340,\phi=0.4, \rho=4\times10^{-5}$ and $c_3$ solved to be $c_3=0.3094$}
    \label{fig:Ekpyrotic_like_energy_density}
\end{subfigure}
\caption{Evolution of anistropic shear (left) and energy density (right) in ekpyrotic-like potential for the simulation shown in Fig. \ref{fig:Ekpyrotic_like_scale_factors}.}
%The initial conditions are $c_1 = 0.57112, c_2 = 0.09828, p_1 = 4130, p_2 =630, p_3 = 2745, \phi=0.19234, \rho=0.00355$ and $c_3$ solved to be $c_3 = 5.54912$}
\label{Ekpyrotic_like_shear_rho}
\end{figure}

The typical behavior of the scale factors, anisotropic shear and energy density in this case are shown in Figs. \ref{fig:Ekpyrotic_like_scale_factors}-\ref{Ekpyrotic_like_shear_rho}. The simulation shown in Fig. \ref{fig:Ekpyrotic_like_scale_factors} depicts two bounces as is apparent from the evolution of the mean scale factor $a$. The ekpyrotic-like potential typically results in multiple bounces in volume since the energy density $\rho$ can become negative and can cause a recollapse in the classical regime if it cancels the contribution of the shear scalar. Here again, all the three directional scalar factors are seen to be increasing with almost the same rate after the last bounce in Fig. \ref{fig:Ekpyrotic_like_scale_factors}, indicating that the universe emerges in a nearly isotropic state after the last bounce due to the ekpyrotic phase induced by the ekpyrotic-like potential. Fig. \ref{Ekpyrotic_like_shear_rho} shows the anisotropic shear and energy density plots in the vicinity of the two bounces. As discussed above, there is a period of slightly negative energy density between the two bounces where the recollapse happens. Additionally, note that the anisotropy is significantly decreased in the second bounce while the energy density is increased resulting in the second bounce being more isotropic, as was seen by the more isotropic behavior of the scale factors after the second bounce for directional scale factors shown in Fig. \ref{fig:Ekpyrotic_like_scale_factors}.

\begin{figure}[H]
    \centering
    \includegraphics[scale=1.1]{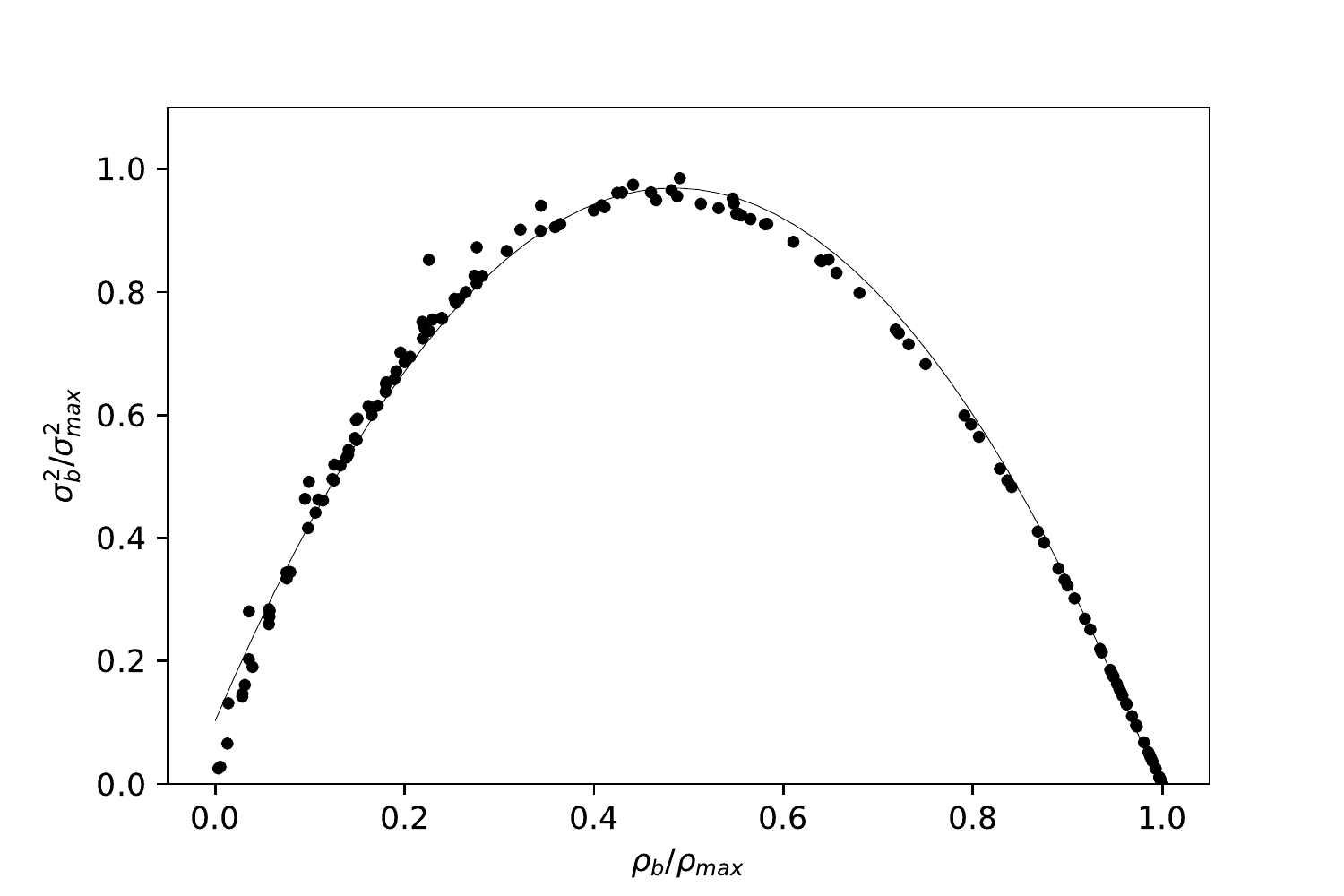}
    \caption{Anisotropic shear versus energy density at the bounce is shown for an ekpyrotic-like potential. Each dot corresponds to one simulation. There are 150 simulations in above plot.}
    \label{fig:Ekpyrotic_like_shear_vs_rho}
\end{figure}

Similar to the previous cases, we can obtain a general understanding of the anisotropic shear and energy density conditions at the time of a bounce with the ekpyrotic-like potential by determining their values at the bounce in a variety of different simulations. The values for $\sigma^2/\sigma_{\mathrm{max}}^2$ versus $\rho / \rho_{\mathrm{max}}$ under an ekpyrotic-like potential at the last bounce are plotted in Fig. \ref{fig:Ekpyrotic_like_shear_vs_rho}, which result in a parabolic shape for this plot analogous to previous cases with a maximum in the middle corresponding to $\sigma_b^2/\sigma_{\mathrm{max}}^2 \approx 1$ and $\rho_b/\rho_{\mathrm{max}} \approx 1/2$. In this case number of bounces turn out to be smaller than the ekpyrotic potential, and results do not significantly change if one considers first or the last bounce in a given time range. As before we consider the last bounce as it serves as a exit to a long phase of the expanding branch.
%We find some deviations from the  with larger deviations from the parabola at smaller values of $\rho$.
The fit parameters in this case for the parabola \eqref{parabola} come out to be $a\approx-3.6936$, $b\approx3.5765$ and $c\approx 0.1033$, which are very close to the values obtained in the cases of the massless scalar field, the $\phi^2$ potential and the ekpyrotic potential discussed above. We also note that the values of the fit parameters depend weakly on the type of potential used. These results are a strong indication that the parabolic relation between $\sigma^2$ and $\rho$ at the bounce may be a feature of the underlying dynamics of the effective Bianchi-I model in LQC.

\section{Towards a generalized effective Friedmann equation}

Extensive numerical simulations reported in the previous section with different types of matter fields which includes a massless scalar field, and a massive scalar field in an inflationary potential, an ekpyrotic and an ekpyrotic-like potential  strongly indicate that energy density and shear scalar have a parabolic relation at the bounce in the effective dynamics of the Bianchi-I model in LQC.  We can use the equation
\begin{equation}\label{parabola1}
  \frac{\sigma_b^2}{\sigma_{\mathrm{max}}^2} = a\left(\frac{\rho_b}{\rho_{\mathrm{max}}}\right)^2 + b \left(\frac{\rho_b}{\rho_{\mathrm{max}}}\right) + c
\end{equation}
to fit the parabolas and Table I summarizes the coefficients of the fits obtained for each of the different matter fields considered in this manuscript. Note that, similar values are obtained for the coefficients $a,b$ and $c$ even when very different types of potentials are used. In other words, the fit parameters seem to depend only weakly on the potential used for the scalar field. We further note that $a$ and $b$ are similar in magnitude, and the vertical intercept $c$ of the parabola is very small and positive.

\begin{table}[H]
\label{Tab:Table}
\centering
\begin{tabular}{|c|c|c|c|}
\hline
\textit{\textbf{Matter content}} & \textbf{a} & \textbf{b} & \textbf{c} \\ \hline
%\textbf{Dust}                                  & -6.23      & 4.90       & 0.078      \\ \hline
%\textbf{Massless Scalar Field}                 & -3.99      & 3.82       & 0.080      \\ \hline
\textbf{Massless scalar field}                 &~ -4.0043~      &~ 3.8148 ~      &~ 0.0812 ~      \\ \hline
%\textbf{$\phi^2$ potential}                 & -3.93      & 3.77       & 0.083      \\ \hline
\textbf{$\phi^2$ potential}                 &~ -3.9279 ~      &~ 3.7739 ~       &~ 0.0828 ~      \\ \hline
%\textbf{Ekpyrotic Potential $V_0=0.01$}                   & -3.61           & 3.52           & 0.088            \\ \hline
\textbf{Ekpyrotic potential $V_0=0.02$}                   &~ -3.7494 ~           &~ 3.6582 ~           &~ 0.0851 ~            \\ \hline
%\textbf{Ekpyrotic-like Potential $u_0=0.0732$}              & -3.71      & 3.60      & 0.094      \\ \hline
\textbf{Ekpyrotic-like potential $u_0=0.0366$}              &~ -3.6936 ~     &~ 3.5765 ~       &~ 0.1033 ~      \\ \hline
%\textbf{Ekpyrotic-like Potential $u_0=0.0183$}              & -3.71      & 3.60       & 0.094      \\ \hline
% \textbf{Ekpyrotic Potential $V_0=0.02$}                   & -3.75           & 3.66           & 0.085            \\ \hline
\hline
\end{tabular}
\caption{Best fit coefficients for the parabola \eqref{parabola1} for different matter fields at the bounce.}

\end{table}

An important question is, what does this parabolic relation at bounce imply for the yet unknown generalized effective Friedmann equation for Bianchi-I spacetime? In the rest of this section, we use our results to draw some  conclusions for the form of the generalized effective Friedmann equation for the effective Bianchi-I spacetime in LQC. Based on our results for the different types of matter fields considered here, we make the assumption that the parabolic relation obtained at the bounce between the energy density and the shear scalar is a general feature of the effective dynamics of the Bianchi-I model. This implies that the generalized effective Friedmann equation must satisfy two requirements. It must reduce to the well known expression \eqref{Friedmann} in the classical limit, while producing the parabolic relation \eqref{parabola1} between $\rho$ and $\sigma^2$ at the bounce in the quantum regime. Since both the expressions \eqref{Friedmann} and \eqref{parabola1} are polynomials in $\rho$ and $\sigma^2$ at the bounce, we work with the assumption that the generalized effective Friedmann equation is a polynomial in both $\rho$ and $\sigma^2$.

First, note that the parabola is obtained at the bounce point, where the mean volume has a turning point and the mean Hubble rate is zero, i.e. the L.H.S. of the generalized Friedmann equation is zero. Since we are in the quantum regime at the bounce where higher order terms in $\rho$ and $\sigma^2$ cannot be ignored, thus, the parabola is either the full effective expression on the R.H.S. of the Friedmann equation or at least a factor of it. For further analysis, we rewrite the parabola as
\begin{equation}\label{parabola2}
    \frac{\sigma_b^2}{\sigma_{\mathrm{max}}^2} + \alpha \left(\frac{\rho_b}{\rho_{\mathrm{max}}}\right)^2 - \beta \left(\frac{\rho_b}{\rho_{\mathrm{max}}}\right) - \eta = 0,
\end{equation}
where $\alpha = |a|, \beta = |b|$ and $\eta = |c|$ are the magnitudes of the fit parameters. Note that the terms linear in $\rho$ and $\sigma^2$ at the bounce are of opposite signs in the parabola. On the other hand, $\rho$ and $\sigma^2$ terms have the same sign in the classical limit as seen from the classical generalized Friedmann equation \eqref{Friedmann}. Thus, the parabola cannot reduce to the classical expression in the classical limit due to the above mentioned sign discrepancy. Also, no other polynomial, which is first order in $\sigma^2_b$ and quadratic in $\rho_b$ will be viable as it will not produce the parabola \eqref{parabola2} in the quantum regime where we cannot ignore the quadratic terms in $\rho_b$. Thus the parabola cannot be the full expression on the R.H.S. of the effective generalized Friedmann equation, and hence this equation must be a polynomial of higher order than the parabola. This implies that the parabola must be a factor of the full expression such  that when the mean Hubble rate vanishes at the bounce a parabolic relation emerges. This motivates the following form,
\begin{equation}\label{Friedmann_eff1}
    H^2=\left(\frac{\sigma^2}{\sigma_{\mathrm{max}}^2} + \alpha \left(\frac{\rho}{\rho_{\mathrm{max}}}\right)^2 - \beta \left(\frac{\rho}{\rho_{\mathrm{max}}}\right) - \eta \right) f(\rho,\sigma^2),
\end{equation}
for the generalized effective Friedmann equation, where $f(\rho,\sigma^2)$ is a polynomial in $\rho$ and $\sigma^2$. This equation produces a parabola in the $\sigma^2$ versus $\rho$ plane at the bounce as observed in our extensive numerical simulations. We now have to determine the form of $f(\rho,\sigma^2)$ so that the equation \eqref{Friedmann_eff1} also reduces to equation \eqref{Friedmann} in the classical limit. The only class of polynomial functions which satisfies these conditions is given by $f(\rho,\sigma^2)=\eta^{-1}(-8\pi G\rho /3 - \sigma^2 /6 + O(\rho^2,\sigma^4))$, causing the generalized effective Friedmann equation to take the form,
\begin{equation}\label{Friedmann_eff2}
    H^2=\left(-\frac{1}{\eta}\right) \left(\frac{\sigma^2}{\sigma_{\mathrm{max}}^2} + \alpha \left(\frac{\rho}{\rho_{\mathrm{max}}}\right)^2 - \beta \left(\frac{\rho}{\rho_{\mathrm{max}}}\right) - \eta \right) \left(\frac{8\pi G\rho}{3} + \frac{\sigma^2}{6} + O(\rho^2,\sigma^4)\right).
\end{equation}
The equation \eqref{Friedmann_eff2} clearly satisfies the two requirements mentioned above. It reduces to the classical expression \eqref{Friedmann} when the energy density and shear scalar are small enough that terms of order quadratic and higher in these can be ignored. In addition, the parabola \eqref{parabola2} provides a set of turning points, i.e. vanishing mean Hubble rate indicating a bounce or recollapse, in the quantum regime where higher order terms in $\rho$ and $\sigma^2$ cannot be ignored.
%Thus, \eqref{Friedmann_eff2} signals a  general polynomial form of the Friedmann equation satisfying these two criteria for the effective Bianchi-I universe.

We conclude this section by noting that further exploration is needed on this issue in different directions. First, one needs to expand the analysis to other matter content such as conventional perfect fluids. One also needs to understand the form of the higher order terms $O(\rho^2,\sigma^4)$ required in $f(\rho,\sigma^2)$ in the generalized effective Friedmann equation. However, the presence of these higher order terms will imply additional bounce/recollapse points in the $\sigma^2$ versus $\rho$ plane apart from the currently known turning points. Apart from the parabolic relation at the bounce, the only other set of turning points presently known for the dynamics of Bianchi-I model are in the classical regime, given by the vanishing of the Hubble rate in equation \eqref{Friedmann}, which gives a negatively sloped straight line $\sigma^2= -16\pi G\rho$ passing through the origin in the $\sigma^2$ versus $\rho$ plane. Thus, information about the higher order terms can be obtained by finding new, as yet unknown turning points in the evolution of volume in the effective LQC dynamics of the Bianchi-I model. Finally, given the complexity of interplay of anisotropic shear and energy density in quantum geometry one must not discount the possibility that the generalized effective Friedmann equation may be sensitive to the explicit form of the equation of state. At least in certain quantizations of isotropic model in LQC this possibility exists \cite{liegener-ps,baofei-ps}.

\section{Conclusions}
As their isotropic counterparts, homogeneous Bianchi-I models are free of the cosmological singularities at the level of effective spacetime description in LQC. Thanks to the underlying quantum geometry,
the energy density and anisotropic shear are universally bounded. The big bang is replaced by a big bounce which generally occurs at a volume much larger than the Planck volume. While in the classical theory, the role of anisotropic shear and energy density on the dynamics as the singularity is approached is extremely well understood, a detailed understanding of the anisotropic nature of the bounce is difficult to obtain, especially for matter with equation of state close to and greater than unity. This is the case for inflationary models as well since the kinetic energy of the inflaton field dominates near the singularity. In anisotropic models in LQC one also lacks a generalized effective Friedmann equation which can provide important insights.
The
physical insights on the interplay of  energy density and anisotropies in the bounce regime have been few and  the way energy density and shear compare with their maximum values at the bounce or any trend was so far unknown.
Understanding the detailed nature of bounce, especially in presence of anisotropies, is also important to construct phenomenologically viable models of a bouncing universe and to understand robust potential signatures in the CMB. Further this issue is important to understand the resolution of singularities in more general models such as Bianchi-II and Bianchi-IX spacetimes and effects on mixmaster dynamics.

  In this manuscript we bring to light a surprising, seemingly universal,  relationship between the anisotropic shear and energy density at the bounce in Bianchi-I models in LQC. Since the energy density only has some potential dominant role for matter content with equation of state $w \approx 1$ and can play an important role for $w > 1$ we consider the cases of a massless scalar field, inflation, and two types of ekpyrotic potentials. Assuming the validity of effective spacetime description we performed extensive numerical simulations with Hamilton's equations with randomized initial conditions. For each case more than 150 simulations were performed.  We find that the values of energy density and the anisotropic shear at the quantum bounce follow a novel parabolic relationship.  This is a strong indication that the parabolic relation is very likely a feature of the effective spacetime of the Bianchi-I model itself. This is further indicated by the fact that the fit parameters of the parabola are found to depend only very slightly on the type of matter field used.  Further investigations with a variety of other matter fields are required to compliment these results.

We note that the existence of the parabolic relationship between energy density and shear scalar at the bounce implies that there is not necessarily a trade-off between them at the bounce. In other words, it is possible that changing initial conditions may lead to both of them increasing or decreasing together at the bounce relative to the previous initial conditions. This is a direct consequence of the parabolic relation, and has been seen in the simulations. Further, as the quantum regime is often dominated by the shear scalar as seen in the simulations, the parabolic trend of the bounce points obtained in the $\sigma^2$ versus $\rho$ plane do not agree with the elliptic relation yielded by the approximate effective Friedmann equation derived in \cite{ChiouKV2007} under assumptions of low shear scalar. A surprising result is that the maximum value of anisotropic shear at the bounce is obtained when energy density at the bounce reaches approximately half of its universal maximum. While the bounce density can be quite close to maximum energy density for small anisotropic shear, our results show that the converse seems difficult. Due to the parabolic nature of relationship which is centered around $\rho_b/\rho_{\mathrm{max}} \approx 1/2$, the anisotropic shear at the bounce never seems to reach a maximum value for small energy density at the bounce at least for the matter content studied in this manuscript.

Due to the complicated form of the equations of motion in the effective spacetime, so far it has not been possible to obtain a generalized effective Friedmann equation which holds in the quantum regime. Using our results at the bounce along with the classical limit, we have described a general form which the effective Friedmann equation may take. However, there are still open questions. It is important to realize that even though the relation obtained at the bounce is simple, the dynamics in the vicinity of the bounce is seen to be much more complicated in our simulations. Thus, using only the results at the exact moment of the bounce and the classical limit, it is not possible to fully grasp the nature of effective dynamics in the whole quantum regime. This is the reason that the higher order terms in our proposed form of the effective Friedmann equation are left unspecified. Knowing the full effective Friedmann equation could lead to an intuitive understanding of the variety of effects seen in the LQC of the Bianchi-I models. We leave it to future investigations to understand the nature of these higher order terms and implications from considering other matter content. Finally, though our results used LQC as the background dynamics they may yield some insights for bouncing anisotropic models in general. It will be interesting to see if this parabolic relation is tied to LQC or is a robust feature of non-singular bouncing anisotropic models.

\section*{Acknowledgments}

We thank Ruotong Zhai for useful discussions. AMM is supported by the REU Site in Physics and Astronomy (NSF Grant No. 1852356) at Louisiana State University.  PS is supported by NSF grant PHY-2110207.

\end{document}